\documentclass[review]{elsarticle}

\usepackage{amssymb}
\usepackage{amsthm}
\usepackage{comment}
\usepackage{color}
\usepackage{graphicx}
\usepackage{upgreek}
\usepackage{hyperref} 
\hypersetup{ colorlinks=true,citecolor=blue, linkcolor=blue, filecolor=blue, urlcolor=blue}
\usepackage{dcolumn}% Align table columns on decimal point
\usepackage{bm}% bold math
\usepackage{amsmath}
\usepackage{amssymb}
\usepackage{scalerel}[2014/03/10]
\usepackage{stackengine}

\usepackage[euler]{textgreek}

\usepackage{multirow}

\newcommand{\tcr}[1]{\textcolor{black}{#1}}

\newcommand{\pd}[2]{\frac{\partial #1}{\partial #2}}

\newcommand{\ie}{\textit{i.e.}}

\begin{document}

\begin{frontmatter}

\title{Data-driven modeling of the aerodynamic deformation and drag for a freely moving drop in the sub-critical Weber number regime}

\author[inst1]{T. Mahmood}

\affiliation[inst1]{organization={Department of Mechanical Engineering, Baylor University},
            city={Waco},
            state={Texas},
            postcode={76798}, 
            country={USA}}

\author[inst2]{A. Tonmoy}
\author[inst2]{C. Sevart}
\author[inst2]{Y. Wang}
\author[inst2]{Y. Ling\corref{cor1}}
\ead{Stanley\_Ling@sc.edu}
\cortext[cor1]{Corresponding author. 
 }

\affiliation[inst2]{organization={Department of Mechanical Engineering, University of South Carolina},            city={Columbia},
            state={South Carolina},
            postcode={29208}, 
            country={USA}}

\begin{abstract}
Accurate prediction of the dynamics and deformation of freely moving drops is crucial for numerous droplet applications. When the Weber number is finite but below a critical value, the drop deviates from its spherical shape and deforms as it is accelerated by the gas stream. Since aerodynamic drag on the drop depends on its shape oscillation, accurately modeling the drop shape evolution is essential for predicting the drop's velocity and position. In this study, 2D axisymmetric interface-resolved simulations were performed to provide a comprehensive dataset for developing a data-driven model. Parametric simulations were conducted by systematically varying the drop diameter and free-stream velocity, achieving wide ranges of Weber and Reynolds numbers. The instantaneous drop shapes obtained in simulations are characterized by spherical harmonics. Temporal data of the drag and modal coefficients are collected from the simulation data to train a {Nonlinear Auto-Regressive models with eXogenous inputs} (NARX) neural network model. The overall model consists of two multi-layer perceptron networks, which predict the modal coefficients and the drop drag, respectively. The drop shape can be reconstructed with the predicted modal coefficients. The model predictions are validated against the simulation data in the testing set, showing excellent agreement for the evolutions of both the drop shape and drag.
\end{abstract}


\end{frontmatter}


%\section{Sample Section Title}
%\label{sec:sample1}
\section{Introduction}
\label{sec:intro}
%Introduction of the problem and motivation
Accurate prediction of the dynamics of freely moving drops is important in numerous droplet applications, such as raindrop impact on aerodynamic surfaces, fuel injection, and spray cooling. Though the interaction between a drop and the surrounding gas flow is complex, the aerodynamic deformation of a drop is typically formulated in an ideal configuration, i.e., an initially stationary and spherical drop is suddenly exposed to an unbounded uniform gas stream \citep{Theofanous_2008a, Marcotte_2019a, Jain_2019a}. In such cases, the drop deformation and dynamics are fully determined by the densities and viscosities of the drop liquid and the gas, $\rho_l$, $\mu_l$, $\rho_g$, and $\mu_g$, the surface tension $\sigma$, the initial drop diameter $D_0$, and the gas stream velocity $U_0$. The subscripts $g$ and $l$ are used to denote the properties for the gas and liquid, respectively, while the subscript $0$ is used to represent the initial state. Neglecting the compressibility \citep{Theofanous_2004a, Theofanous_2007a, Sharma_2021c} and non-Newtonian \citep{Joseph_1999a,Theofanous_2013a} effects, the drop shape deformation and dynamics can be fully characterized by four independent dimensionless parameters: the Weber number $\textit{We} = \rho_g U_0^2 D_0/\sigma$, the Reynolds number $\textit{Re} = \rho_g U_0 D_0/\mu_g$, the Ohnesorge number $\textit{Oh} = \mu_l/\sqrt{\rho_l D_0 \sigma}$, and the gas-to-liquid density ratio $\textsl{r}=\rho_g/\rho_l$ \citep{Pilch_1987a, Hsiang_1992a, Joseph_1999a, Guildenbecher_2009a}. Except for high-pressure applications \cite{Manin_2014a}, the density ratio between gas and liquids is generally small and the effect of $\textsl{r}$ is secondary \cite{Marcotte_2019a}. The Ohnesorge number $\textit{Oh}$ measures the relative importance of liquid viscosity compared to surface tension. For low-viscosity liquids like water, $\textit{Oh}$ is generally small except for very small drops, and therefore surface tension is the dominant force to resist drop deformation or breakup. Previous studies showed that the effect of $\textit{Oh}$ on \tcr{the criteria for the onset of breakup} is small when $\textit{Oh}<0.1$ \cite{Hsiang_1992a}. The Weber number \textit{We} is generally used to characterize drop breakup criteria and breakup modes \cite{Hsiang_1992a}. For water drops in air, the critical Weber number is about 11, under which a drop will only experience shape oscillation but will not break. The Reynolds number \textit{Re} dictates the \tcr{viscous gas flows in outside the drop}, such as the boundary layer separation and the wake structure. For a spherical particle, the wake becomes turbulent when $\textit{Re}>3000$ \cite{Yun_2006a,Tiwari_2020b}.

In the sub-critical Weber number regime, the drop experiences shape oscillation when it is accelerated by the aerodynamic drag. The complex interaction between the shape oscillation and the surrounding flow makes accurate prediction of the drop shape and drag challenging. The oscillation dynamics for drops with finite \textit{We} is more complicated than capillary oscillation of a drop in a quiescent environment \cite{Rayleigh_1879a, Lamb_1932a, Reid_1960a, Miller_1968a, Prosperetti_1980a}. As the drop experiences a large-amplitude oscillation, the nonlinear effect will modulate the drop oscillation dynamics \cite{Tsamopoulos_1983a, Basaran_1992a}, e.g., the drop oscillation frequency decreases as the oscillation amplitude increases. Furthermore, the drop oscillation amplitude decreases over time due to viscous dissipation, so the nonlinear effect will be reduced as time elapses and as a result, the drop oscillation frequency will change over time.  As the external gas flow and the liquid flow inside the drop can be significantly modulated by the drop shape oscillations \cite{Bergeles_2018a, Lalanne_2013a, Zhang_2019b},  the drag coefficient for a deforming drop will be significantly different from a spherical drop in the zero-\textit{We} limit, and will evolve in time in an oscillatory manner. How to incorporate the effect of shape oscillation in the drop drag model remains an open question \cite{Lin_2022a}.

In practical simulations of sprays consisting of a large number of droplets, it is impractical to resolve the interface of each individual drop. Instead, the drops are represented by point particles and traced in a Lagrangian framework \citep{Apte_2003a, Pai_2006a, Balachandar_2009a}. The drop drag and shape deformation, along with other physics like heat and mass transfer between the drop and the surrounding flow \cite{Abramzon_1989a, Boyd_2023s, Boyd_2023c} and aerodynamic breakup, need to be represented by subgrid models \citep{ORourke_1987a, Hsiang_1992a}. For drops with finite $\textit{We}$, traditional drag models for spherical drops \cite{Maxey_1983a} in the zero-$\textit{We}$ limit will be invalid. Though different physics-based models have been proposed \citep{ORourke_1987a, Kulkarni_2012a, Rimbert_2020o, Jackiw_2021a}, significant discrepancies between the model predictions and high-fidelity simulation and experiment were observed \cite{Ling_2023a}. The fundamental challenge of physics-based models lies in the necessary assumptions/simplifications of the drop shape and surrounding flows, such as assuming the drop shape is a spheroid, which are typically valid only for a limited time period and small $\textit{We}$. 

Data-driven modeling is an important alternative to develop sub-scale models for multi-phase flows, and its capability is significantly enhanced by the rapid development of machine-learning techniques. Wan et al. \cite{Wan_2020a} presented a data-driven kinematic model to predict the trajectories of bubbles in high-\textit{Re} fluid flow using a recurrent neural network consisting of long short-term memory (LSTM) layers. Studies have been dedicated to developing machine-learning models of quasi-steady drag for particle-laden flow with finite particle volume fractions \cite{He_2019b, Siddani_2021a, Siddani_2021b, Siddani_2023y}. To the knowledge of the authors, there are not yet machine-learning models for the drag force acting on deforming drops. The additional time-dependent shape deformation of drops makes the modeling more challenging compared to the particle counterpart.

The goal of the present study is to develop a data-driven model to predict the time evolutions of drop shape and drag for a freely-moving drop in the sub-critical Weber number regime. Though it is of highly interest to model a wider range of parameters that cover also the aerodynamic breakup of drops, the present study will be focused on the aerodynamic deformation of drops that will not experience \tcr{breakup}. To provide the data to train and to test the model, interface-resolved simulations using volume-of-fluid (VOF) method will be performed. The Nonlinear Auto-Regressive models with eXogenous inputs (NARX) recurrent neural network  \cite{Lin_1995a, Lin_1996a} will be used to develop the data-driven models. The rest of the paper will be organized as follows. We will first define the problem and the parameter space of interest in section \ref{sec:problem_description}. The numerical methods and solver for the interface-resolved simulations will be presented in section \ref{sec:simulation_methods}, followed by the simulation results shown in section \ref{sec:sim_results}. The machine-learning model architecture will be described in section \ref{sec:model}, and the model predictions and comparison with simulation data will be presented in section \ref{sec:model_results}. Finally, we will conclude key findings of the present study in section \ref{sec:conclusions}.

\section{Problem description and parameter space}
\label{sec:problem_description}
\subsection{Parameter space of interest}
In the present study, we consider that an initially stationary and spherical drop is suddenly exposed to a uniform gas stream \cite{Ranger_1969a, Hsiang_1992a, Tang_2023a, Ling_2023a}. The physical parameters and the dimensionless parameters are listed in Tables \ref{tab:Osc_phy_param} and \ref{tab:Osc_nodim_param}. The drop liquid and gas are taken to be water and air, respectively. As a result, $\textsl{r}=0.0012$ is fixed and for such a low value the density ratio effect is negligible \cite{Marcotte_2019a}. A parametric study is performed by varying $U_0$ and $D_0$. \tcr{As a result, $\textit{We}$, $\textit{Oh}$ and $\textit{Re}$ vary from case to case.} For the ranges of $U_0$ and $D_0$ considered, $\textit{Oh}\le 0.086$, \tcr{and it is expected that the effect of liquid viscosity, though is present, is less important compared to surface tension. Previous studies indicated that the impact of $\textit{Oh}$ on the critical Weber number $\textit{We}_{cr}$ is negligible for the present ranges of $\textsl{r}$ and $\textit{Oh}$, and $\textit{We}_{cr}\approx 11\pm 2$ \cite{Hsiang_1992a}}. \tcr{Futhermore, the drop drag mainly depends on $\textit{Re}$,} therefore, $\textit{We}$ and $\textit{Re}$ are the key controlling parameters. The ranges of which considered in the present study are : $0.1\leq\textit{We}\leq 10$ and $10\leq \textit{Re}\leq1000$. The focus is on the sub-critical regime, $\textit{We}<\textit{We}_{cr}$, so that the drop will only undergo oscillation but will not break and form child droplets. Furthermore, we have only considered moderate \textit{Re} so that the less expensive 2D axisymmetric simulations will remain good approximation, and neglecting the 3D flow features in the drop wake will not lead to significant effect on the drop deformation and drag  \cite{Boyd_2023c}. The low computational costs for 2D axisymmetric simulations will allow us to consider a large number cases.  To guarantee the compressibility effect is negligible, it is taken that $\textit{M}<0.3$, which leads to a constraint that $\textit{Re}>38.87\textit{We}$. Furthermore, we consider the drop diameter to be smaller than 10 mm, which then \tcr{leads} to another boundary in the $\textit{Re}$-$\textit{We}$ space, \ie, $\textit{Re}<1632\sqrt{\textit{We}}$. The final parameter space of interest is shown in Fig.~\ref{fig:osc_setup}(b). 

\begin{table}
\setlength{\tabcolsep}{3.5pt}
  \caption{Fluid properties for simulation cases.}
\centering
  \begin{tabular}{cccccccccc} 
  \hline
 $\rho_{l}$      & $\rho_{g}$     & $\mu_{l}$ & $\mu_{g}$ & $\sigma$ & $D_0$ & $U_0$\\ 
    (kg/m$^{3}$) & (kg/m$^{3}$) & (Pa\,s) & (Pa\,s)  & (N/m)	& (m)	& (m/s)\\
    \hline
$1000$             & $1.2$                  & $0.001$ & $0.000018$ & $0.072$  & $5.8\times10^{-3} - 5\times10^{-6}$ & $1.34 - 88.02$ \\ 
\hline
  \end{tabular}
  \label{tab:Osc_phy_param}
\vspace{-4mm}
\end{table}

\begin{table}
\setlength{\tabcolsep}{3.5pt}
  \caption{Range of dimensionless parameters for simulation cases.}
  \centering
  \begin{tabular}{cccccccccc} 
  \hline
  \textit{We}  &\textit{Re} &\textit{Oh}	&\textsl{r}\\ 
  $\rho_{g}U_0^{2}D_0/\sigma$ &$\rho_{g}U_0D_0/\mu_{g}$ &$\mu_{l}/\sqrt{\rho_{l}D_0\sigma}$ &$\rho_g/\rho_l$\\
  \hline
  $0.1 - 10$ & $10-1000$ & $0.0012 - 0.0861$ &$0.0012$ \\ 
  \hline
  \end{tabular}
  \label{tab:Osc_nodim_param}
\vspace{-4mm}
\end{table}

\subsection{Cases of study}
\begin{figure}
\centering
\includegraphics[trim={0cm 0cm 0cm 0cm},clip,width=0.95\textwidth]{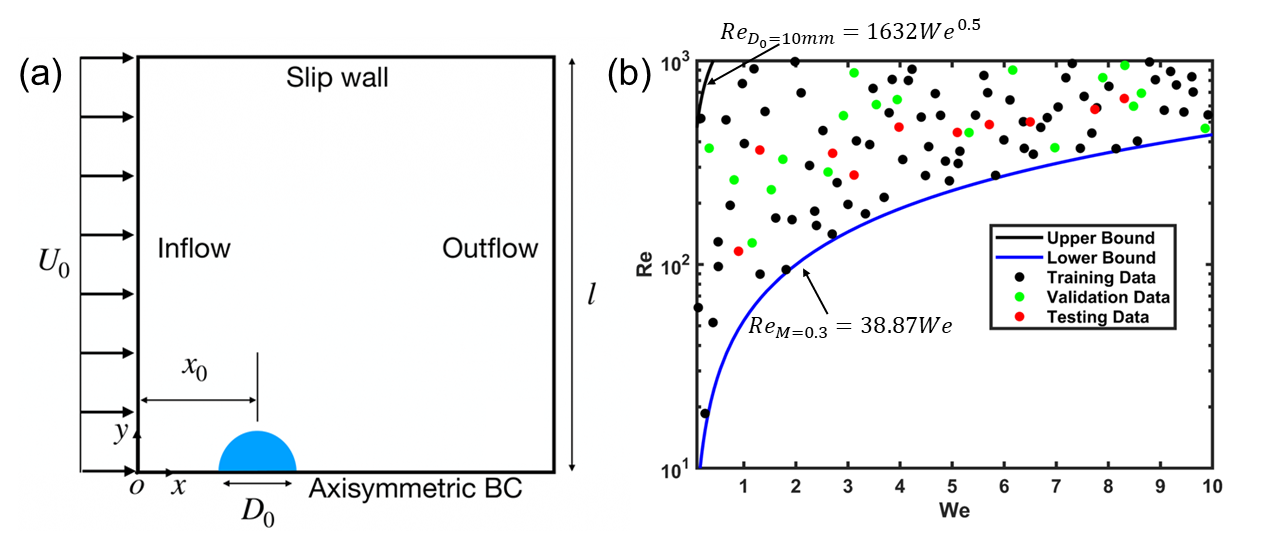}
\caption{(a) Schematic of simulation domain and (b) parameter space of interest in the \textit{We}-\textit{Re} plane, where \textit{We} and \textit{Re} are in linear and log scales, respectively.}
\label{fig:osc_setup}
\vspace{-4mm}
\end{figure}

A total of 102 cases were selected by performing a Latin Hypercube Sampling using the maximum-minimum distance criteria over the parameter space of interest, see Fig. \ref{fig:osc_setup}(b)). Among them, 92 cases will be used for data-driven model development, and are further split randomly into training and validation sets by an 80:20 ratio. Training sets are used to train the model and validation sets will ensure the model's robustness by preventing the over-fitting of the training data. The remaining 10 cases will be used as the testing set which will provide the unbiased evaluation of the trained model. \tcr{The simulation cases used as training, validation, and testing datasets are detailed in the \ref{sec:dataset}.}

\section{Interface-resolved simulations}
\label{sec:simulation_methods}
\subsection{Simulation methods}
Interface-resolved simulations were performed for the selected cases to generate the data to train and test the data-driven model. The two-phase interfacial flows are governed by the incompressible Navier-Stokes equations with surface tension,
\begin{align}
	\rho \left(\pd{ u_i}{t} + u_i \pd{u_j}{x_j}\right) & = -\pd{p}{x_i} + \pd{}{x_j}\left[ \mu\left( \pd{u_i}{x_j} + \pd{u_j}{x_i} \right) \right] + \sigma \kappa\delta_s n_i\, ,	
	\label{eq:mom}\\
	\pd{u_i}{x_i} & =0\, .	
	\label{eq:cont}
\end{align}
where $\rho, u_i, p, \mu$ represent density, velocity, pressure and viscosity, respectively.  The Dirac distribution function $\delta_s$ is localized on the interface. The surface tension coefficient is denoted by $\sigma$, while $\kappa$ and $n_i$ represent the curvature and normal vector of the interface. 

The gas and liquid phases are distinguished by the liquid volume fraction $c$, the evolution of which follows the advection equation: 
\begin{align}
  \pd{c}{t} + u_i \pd{c }{x_i} & = 0 \, .
  \label{eq:adv}
\end{align}
After spatial discretization, the cells with pure liquid or gas will exhibit $c=1$ and 0, respectively, while for cells containing the gas-liquid interface, $c$ is a fractional number. The density and viscosity are both defined based on the arithmetic mean
\begin{align}
  \rho & = \rho_l c + \rho_g (1-c)  \, , \\
  \mu & = \mu_l c + \mu_g (1-c)\, .
  \label{eq:adv}
\end{align}

The present simulations are conducted using the open-source solver \textit{Basilisk}. The governing equations are solved using a finite-volume method. The projection method is used to incorporate the incompressibility condition. Sharp interfaces separating the two phases are traced by solving the advection equation via a mass-momentum consistent geometric Volume-of-Fluid (VOF) method \citep{Arrufat_2020a, Zhang_2020a}. The balanced-force method is used to discretize the singular surface tension term in the momentum equation \citep{Popinet_2009a}. The interface curvature required to calculate surface tension is computed based on the height-function (HF) method \citep{Popinet_2009a}. The staggered-in-time discretization of the volume fraction/density and pressure leads to a formally second-order-accurate time discretization \citep{Popinet_2009a}. An quadtree mesh is used to discretize the 2D computational domain, which allows for adaptive mesh refinement (AMR) in user-defined regions. The mesh adaptation is based on the wavelet estimate of the discretization errors of specified variables \citep{Hooft_2018a}. Validation of the numerical methods and the solver \textit{Basilisk} in resolving drop deformation and breakup can be found in our previous studies \cite{Zhang_2019b, Sakakeeny_2020a, Sakakeeny_2021a, Sakakeeny_2021b, Ling_2023a, Boyd_2023c, Boyd_2023s}. 

\subsection{Simulation setup}
\label{sec:Osc_setup}
The computational domain is shown in Fig. \ref{fig:osc_setup}(a). The velocity of the drop and the gas in the domain is initially zero. The velocity boundary condition is invoked on the left boundary of the domain, with a pressure outflow boundary condition invoked on the right boundary. Due to the incompressibility condition, the gas is suddenly accelerated to $U_0$ in an infinitesimal time (one time step in the simulation). The bottom is the axis and the top is a slip wall. The computational domain has an edge length of $l=64D_0$, and the drop is initially placed $x_0=3D_0$ away from the left boundary. The computational domain is discretized by a quadtree mesh, which is dynamically adapted based on the wavelet estimates of the discretization errors of the liquid volume fraction and the velocity components. The minimum cell size $\Delta$ is controlled by the maximum refinement level $\mathcal{L}$, i.e., $\Delta=l/2^\mathcal{L}$. In the present study, $\mathcal{L}=13$ is used, corresponding to 128 minimum quadtree cells across the initial drop diameter, i.e., $N=D_0/\Delta=128$. The grid-refinement study, to be discussed in section \ref{sec:Osc_verify}, confirms that the mesh resolution is sufficient.

\section{Simulation results and data processing}
\label{sec:sim_results}
\subsection{General behavior}
\label{sec:Osc_drop_dynamics}
\begin{figure}
\centering
\includegraphics[trim={0cm 0cm 0cm 0cm},clip,width=0.60\textwidth]{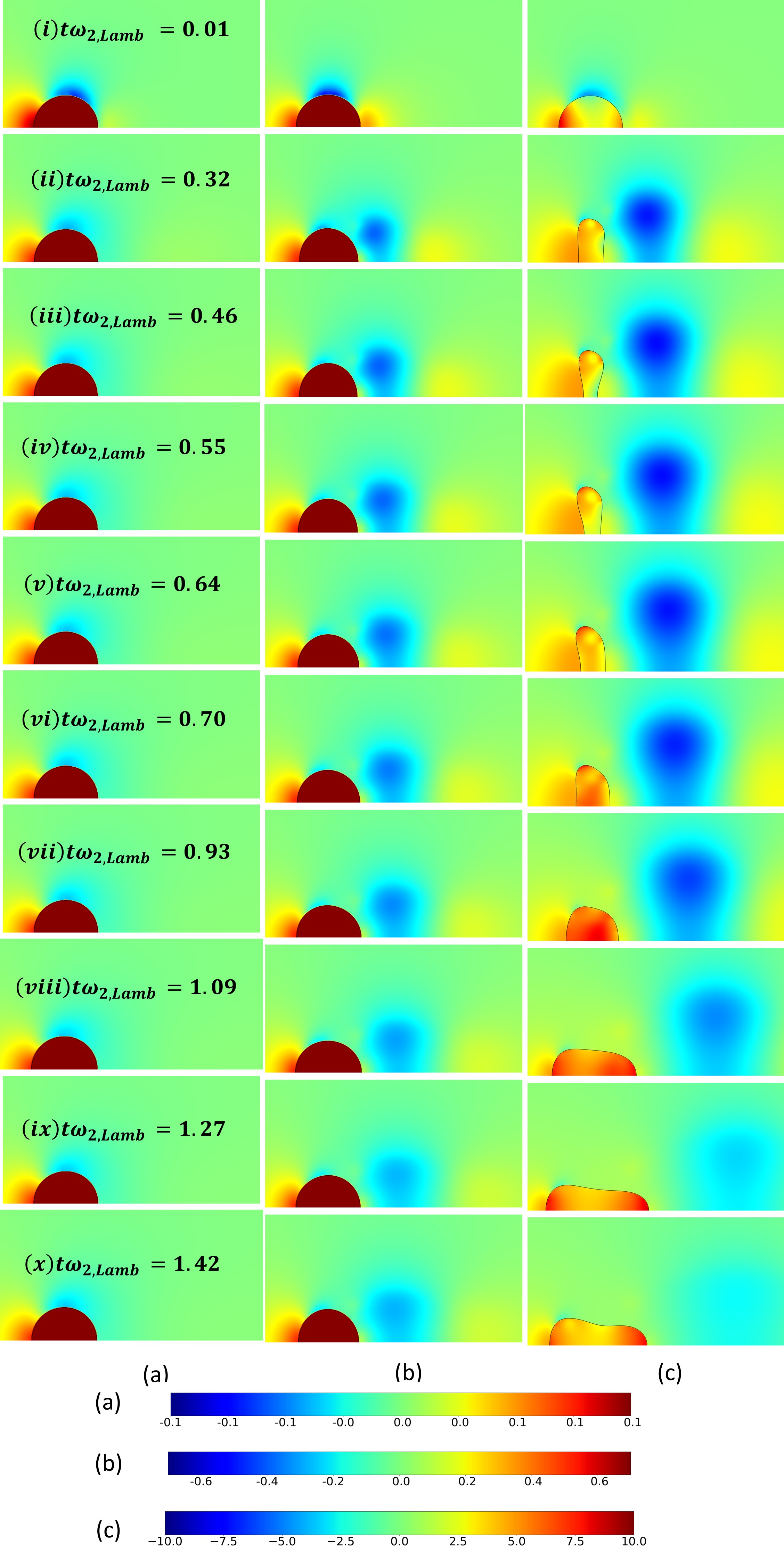}
\caption{Temporal evolutions of the pressure fields for three representative cases with different \textit{We} and \textit{Re}. (a)$\textit{We} = 0.13$, $\textit{Re} = 61.42$, (b) $\textit{We} = 0.97$, $\textit{Re} = 772.89$, (c) $\textit{We} = 9.60$, $\textit{Re} = 834.65$. }
\label{fig:osc_pre_3c}
\vspace{-4mm}
\end{figure}

The time evolution of the pressure fields and drop surface for different $\textit{We}$ and $\textit{Re}$ are shown in Fig.~\ref{fig:osc_pre_3c}. The cases in (a), (b), and (c) are for $(\textit{We},\textit{Re})=(0.13,61.42)$, $(0.97,772.89)$, and $(9.60,834.65)$, which represent low-\textit{We}-low-\textit{Re}, low-\textit{We}-high-\textit{Re}, and high-\textit{We}-high-\textit{Re} regimes, respectively. The high stagnation pressure near the windward and leeward poles of the drop drives radial flow towards the periphery, leading to the flattening of the drop in the streamwise direction \cite{Villermaux_2009a, Jackiw_2021a}. \tcr{The corresponding $\textit{Oh}$ for cases (a), (b), and (c) are 0.011, 0.0025, and 0.0071, respectively.} Due to the low $\textit{Oh}$, surface tension is the dominant force resisting drop deformation. It is evident that cases with low $\textit{We}$ show only mild deformation throughout the process, whereas, for the case with high $\textit{We}$, the deformation is more pronounced, and the drop transitions from a sphere to a flat disk, as seen in Fig. \ref{fig:osc_pre_3c}(c). Since $\textit{We}$ is below the critical value, surface tension is sufficient to revert the drop to an elongated shape, causing it to deform in an oscillatory manner. Although cases (a) and (b) have low $\textit{We}$ and thus similar drop shapes, their $\textit{Re}$ values are markedly different, leading to distinct wake structures and consequently different drag forces on the drop. 

\subsection{Drop shape characterization}
\label{sec:osc_shape_char}
To characterize the shape of the drop, we employ the spherical coordinate system, and the radius on the drop surface can be expressed as a function of the colatitude $\theta$ and time $t$, i.e., $R=R(\theta ,t)$. The instantaneous drop shape can be decomposed into axisymmetric spherical harmonic modes, represented by Legendre polynomials $P_n$
\begin{align}
\frac{R(\theta, t)-R_0}{R_0}=\sum_{n=0}^{\infty} C_{n}(t)P_{n}(\cos(\theta))\, ,
\label{eq:osc_sp_harmonic}
\end{align}
where $R_0=D_0/2$ is the volume-based radius, $C_n$ represents the coefficient for the mode $n$, which varies over time. For the current problem, keeping modes with $n\le10$ is sufficient to accurately represent the drop shape for all cases and all time. 
Given the instantaneous shape of the drop, $C_n$ can be computed by the Legendre-Fourier transform 
\begin{align}
C_{n}(t)= \frac{2n+1}{2}\int_{-1}^{1} \frac{R(\theta,t)-R_0}{R_0}P_{n}(\cos(\theta)) d(\cos(\theta)).
\label{eq:osc_mode_coeff}
\end{align}

\begin{figure}
\centering
\includegraphics[trim={0cm .8cm 0cm 0.8cm},clip,width=0.90\textwidth]{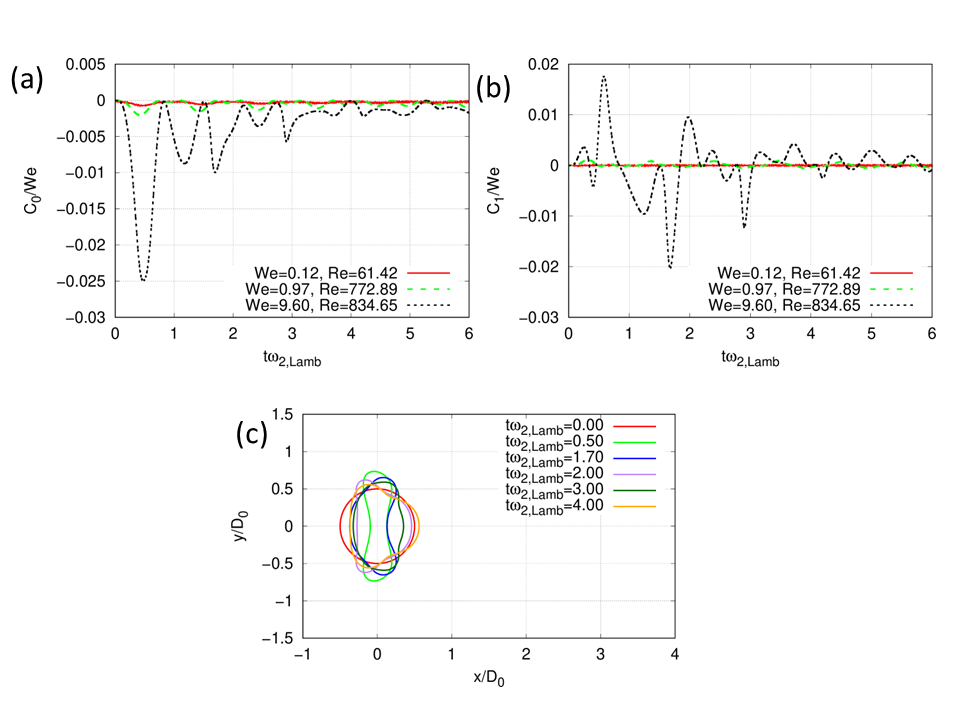}
\caption{Temporal evolutions of (a) $C_0$ and (b) $C_1$ for three representative cases with different \textit{We} and \textit{Re}. (c) Time snapshots of the drop shapes for the case $\textit{We}=9.60$ and $\textit{Re}=834.65$. }
\label{fig:osc_C0_not_zero}
\vspace{-4mm}
\end{figure}

When the drop deformation amplitude is small, $C_0=0$ if the volume is constant and $C_1=0$ if $R(\theta,t)$ is defined based on the drop center. However, when the drop deformation amplitude is high, as for cases with high \textit{We}, $C_0$ and $C_1$ are not identical to zero when the centroid is used as the origin of the coordinate system, see Figs.~\ref{fig:osc_C0_not_zero}(a) and (b). The deviations from zero are more profound for the case $\textit{We}=9.60$, for which the drop deforms more significantly, which can be observed from the snapshots of the drop shapes shown in Fig.~\ref{fig:osc_C0_not_zero}(c). As the deformation amplitude decreases over time, the magnitudes for $C_0$ and $C_1$ also reduce. The results here indicate that it is necessary to include $C_0$ and $C_1$ for shape characterization.

\subsection{Grid refinement study}
\label{sec:Osc_verify}
To verify that the mesh resolution is sufficient to resolve the drop shape oscillation, a grid-refinement study is performed using the same representative cases shown in Fig.~\ref{fig:osc_pre_3c}. Since these cases represent the corners in the parameter space of interest, convergence of results for these cases will guarantee the mesh is fine enough for other cases considered. 

\begin{figure}
\centering
\includegraphics[trim={0cm 0.6cm 0.cm 0.6cm},clip,width=0.90\textwidth]{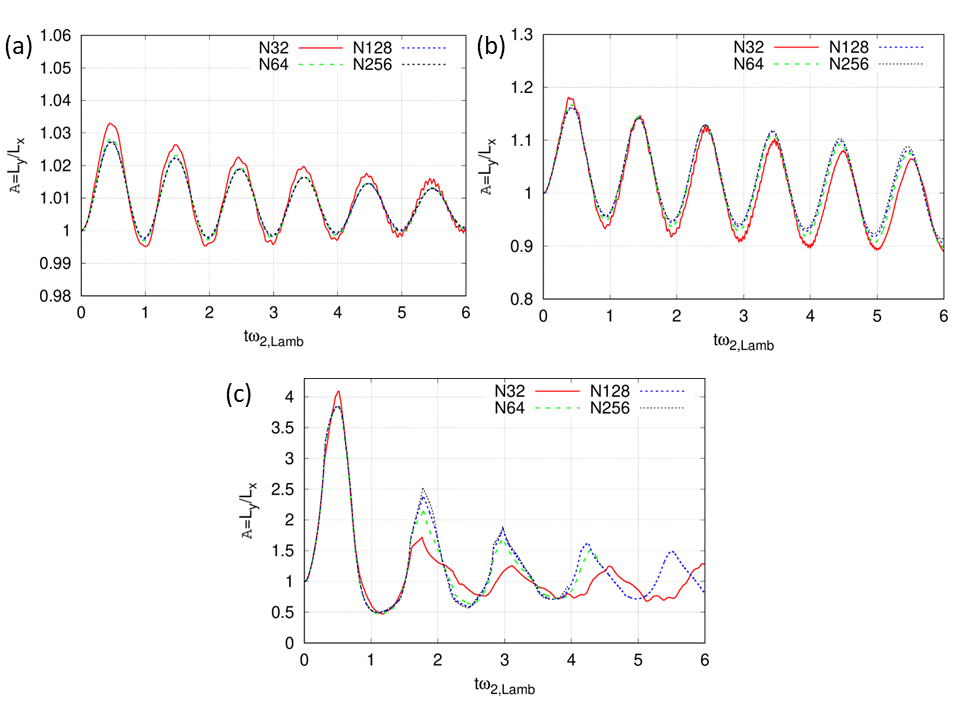}
\caption{Temporal evolutions of the aspect ratios (${A}=L_{y}/L_{x}$) for different levels of mesh refinement $\textit{N}=D_0/\Delta_{min}=32-256$ for for three representative cases with different \textit{We} and \textit{Re}. (a)$\textit{We} = 0.13$, $\textit{Re} = 61.42$, (b) $\textit{We} = 0.97$, $\textit{Re} = 772.89$, (c) $\textit{We} = 9.60$, $\textit{Re} = 834.65$.
}
\label{fig:osc_verify}
\vspace{-4mm}
\end{figure}

The time evolutions of the aspect ratio, ${A}=L_y/L_x$, where $L_x$ and $L_y$ are the streamwise and lateral widths of the drop, for different cases and mesh resolutions, are shown in Fig.~\ref{fig:osc_verify}. Four different mesh refinement levels have been tested, $\mathcal{L}=11$, $12$, $13$, and $14$, which correspond to $N=D_0/\Delta=32$, $64$, $128$, and $256$, respectively. Time has been normalized with the Lamb frequencies of the dominant second mode $\omega_{2,\mathrm{lamb}}$. The Lamb frequency of the $n^{th}$ axisymmetric mode is given as \cite{Lamb_1932a}
\begin{align}
    \omega_{n,Lamb}=\sqrt{\frac{(n-1)n(n+1)(n+2)\sigma}{[(n+1)\rho_l+n\rho_g]R_0^3}}.
    \label{eq:Osc_lamb}
\end{align}
The results for $N=128$ and 256 match very well, indicating the mesh $N=128$ is sufficient to yield converged results, which is used for the parametric simulations.

\subsection{Temporal evolutions of modal coefficients}
\begin{figure}
\centering
\includegraphics[trim={0cm 0cm 0cm 0cm},clip,width=1.\textwidth]{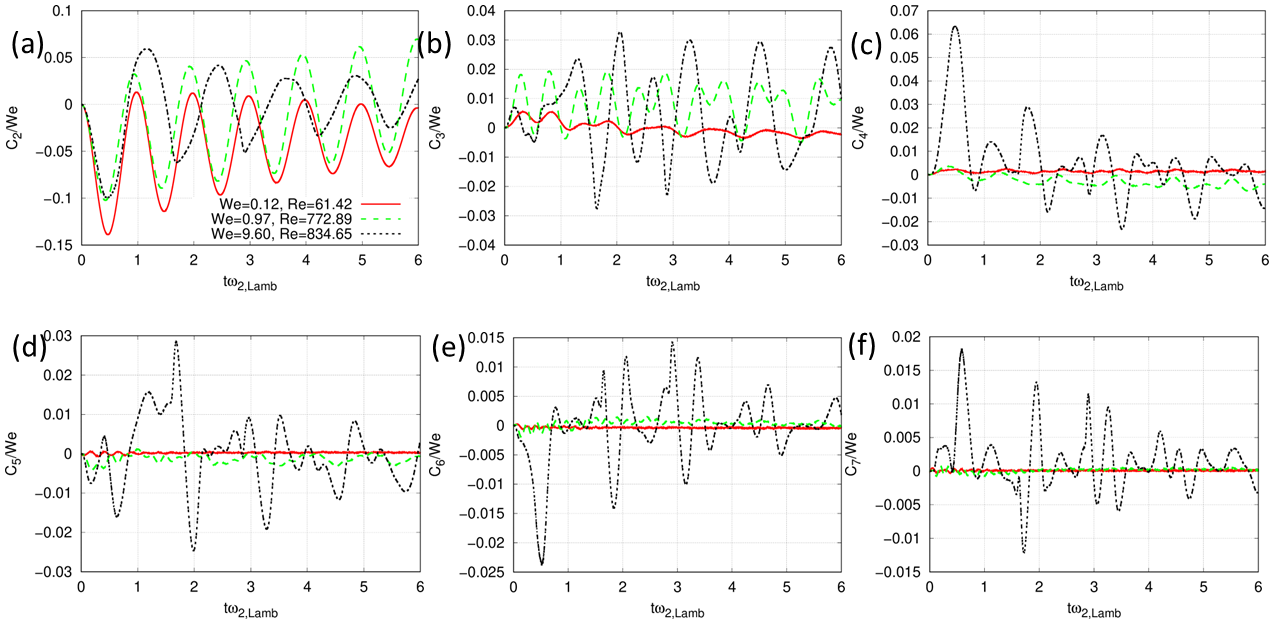}
\caption{Temporal evolutions of the modal coefficients $C_2$ to $C_7$ for three representative cases.}
\label{fig:osc_C_t}
\vspace{-4mm}
\end{figure}

The temporal evolutions of the Fourier-Legendre coefficients for modes $n=2-7$ for different \textit{We} and \textit{Re} are plotted in Fig.~\ref{fig:osc_C_t}. As the amplitude of the modal coefficient variation increases with \textit{We}, we have plotted the ratio of $C_n$ to \textit{We} instead. The differences in the results between different cases are reduced for $C_n/\textit{We}$, which also helps to improve the accuracy of the data-driven model.

All the modal coefficients are initially zero since the drop is spherical at $t=0$. As the drop starts to deform, $C_n$ varies over time. It is shown that the oscillation of $C_2$ exhibits a single frequency. For low \textit{We} cases, the frequency is close to that of the 2nd Lamb mode, while a lower frequency is observed for the high $\textit{We}=9.60$ case. Furthermore, the equilibrium state of the $C_2$ oscillation is not zero, indicating the equilibrium shape is not a perfect sphere. The time evolutions of the coefficients for modes $n>2$ are more complicated, in particular for the high $\textit{We}$ case, due to the non-linear effects \cite{Becker_1991a, Basaran_1992a, Zhang_2019b}. It is difficult to model the time evolutions of each mode using simple explicit functions, as can be done for small-amplitude drop oscillations \cite{Zhang_2019b}.

%\begin{figure}
%\centering
%\includegraphics[trim={0cm 0cm 0cm 0cm},clip,width=0.90\textwidth]{osc_fft_three_cases.png}
%\caption{Frequency spectra of Fourier-Legendre coefficients indicating mode coupling. Amplitude is in logarithmic scale (a)$\textit{We}=0.13$, $\textit{Re}=61.42$, (b)$\textit{We}=0.97$, $\textit{Re}=772.89$ and (c)$\textit{We}=9.6$, $\textit{Re}=834.651$)}
%\label{fig:osc_freq_spec}
%\vspace{-4mm}
%\end{figure}

\subsection{Temporal evolutions of drag coefficient}
\label{sec:evo_drag}
The aerodynamic drag causes the drop to accelerate along the streamwise direction. The mean x-velocity of the drop is calculated by the integration of the velocity and VOF fields, 
\begin{align}
u_{d} = \frac{\int {c\,u\, dV}}{\int{c\, dV}}
\label{eq:vel_def}
\end{align}
based on which the drop acceleration can be calculated as $du_d/dt$, and the drag coefficient is evaluated as 
\begin{align}
    C_{D}= \frac{2m_d}{\rho_g (U_0-u_d)^2 \pi R^2} \frac{du_d}{dt}
    \label{eq: drag_def}
\end{align}
where $m_d$ is the mass of the drop. Here, $C_D$ is defined based on the instantaneous relative velocity $(U_0 - u_d)$, and the drop frontal area $(\pi R^2)$ estimated by the lateral radius $R$. We normalize the drag coefficient with the standard drag correlation for spherical particle corresponding to \textit{Re}, 
\begin{align}
    C_{D,std}=\frac{24}{\textit{Re}} (1+0.15\textit{Re}^{0.687}) + \frac{0.42}{1+\frac{42500}{\textit{Re}^{1.16}}}
\label{eq:Drag_Co_sp}
\end{align}

\begin{figure}
\centering
\includegraphics[trim={0cm 0cm 0cm 0cm},clip,width=0.5\textwidth]{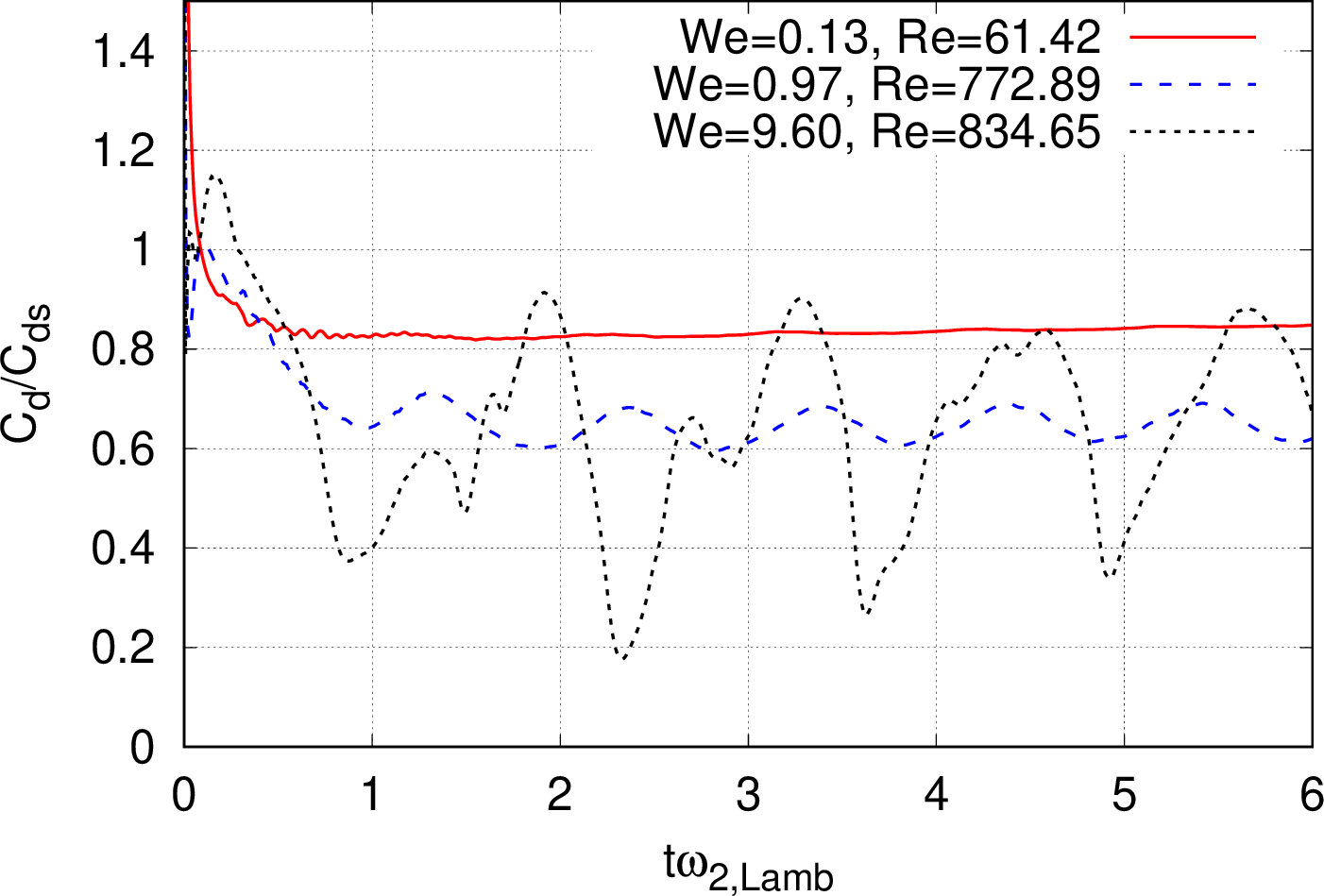}
\caption{Temporal evolutions of drag coefficients for three representative cases with different \textit{We} and \textit{Re}. }
\label{fig:osc_drag_evo}
\end{figure}

The time evolutions of $C_D$ for different \textit{We} and \textit{Re} are shown in Fig.~\ref{fig:osc_drag_evo}. \tcr{The three cases shown are the same cases shown in Figs.~\ref{fig:osc_pre_3c} and \ref{fig:osc_verify}.}
Initially $C_D$ is much higher than $C_{D,std}$ due to the impulsive acceleration at time zero and the resulting unsteady forces (added-mass and Basset history force) \cite{Ling_2013b}. The drag coefficient will approach the steady drag after the transition phase, \tcr{the duration of which is} dictated by the viscous unsteady time scale. For the case (a) with low $\textit{We}=0.13$ and $\textit{Re}=61.42$, it is seen that $C_D/C_{D,std}$ reaches a plateau, about 0.82, see Fig.~\ref{fig:osc_drag_evo}. \tcr{In contrast, $C_D$ for case (b) oscillates in time and the oscillation frequency matches with that for the shape oscillation. The increase in $C_D$ oscillation amplitude is attributed to the  increase of $\textit{We}$ and the resulting stronger drop shape deformation, see Fig.~\ref{fig:osc_verify}. Compared to case (a), the wake structure for case (b) also varies as $\textit{Re}$ increases significantly, see Fig.~\ref{fig:osc_pre_3c}, which contributes to the lower plateau value of $C_D$.} For the case (c) with high $\textit{We}=9.60$, $C_D$ exhibits a more complex oscillation with multiple frequencies and larger amplitude\tcr{, which is due to the large-amplitude multi-mode shape oscillations, see Fig.~\ref{fig:osc_verify}.} \tcr{The oscillation in $C_D$ with a frequency similar to the shape oscillation clearly demonstrates that the drag and shape evolutions are closely coupled. Therefore, it is necessary to incorporate the drop shape evolution into the data-driven model to accurately predict the drag.}

\subsection{Effect of \textit{We} and \textit{Re} on aspect ratio and dominant frequency}
\label{sec:effect_We_Re}
\begin{figure}
\centering
\includegraphics[trim={0cm 0cm 0cm 0cm},clip,width=0.95\textwidth]{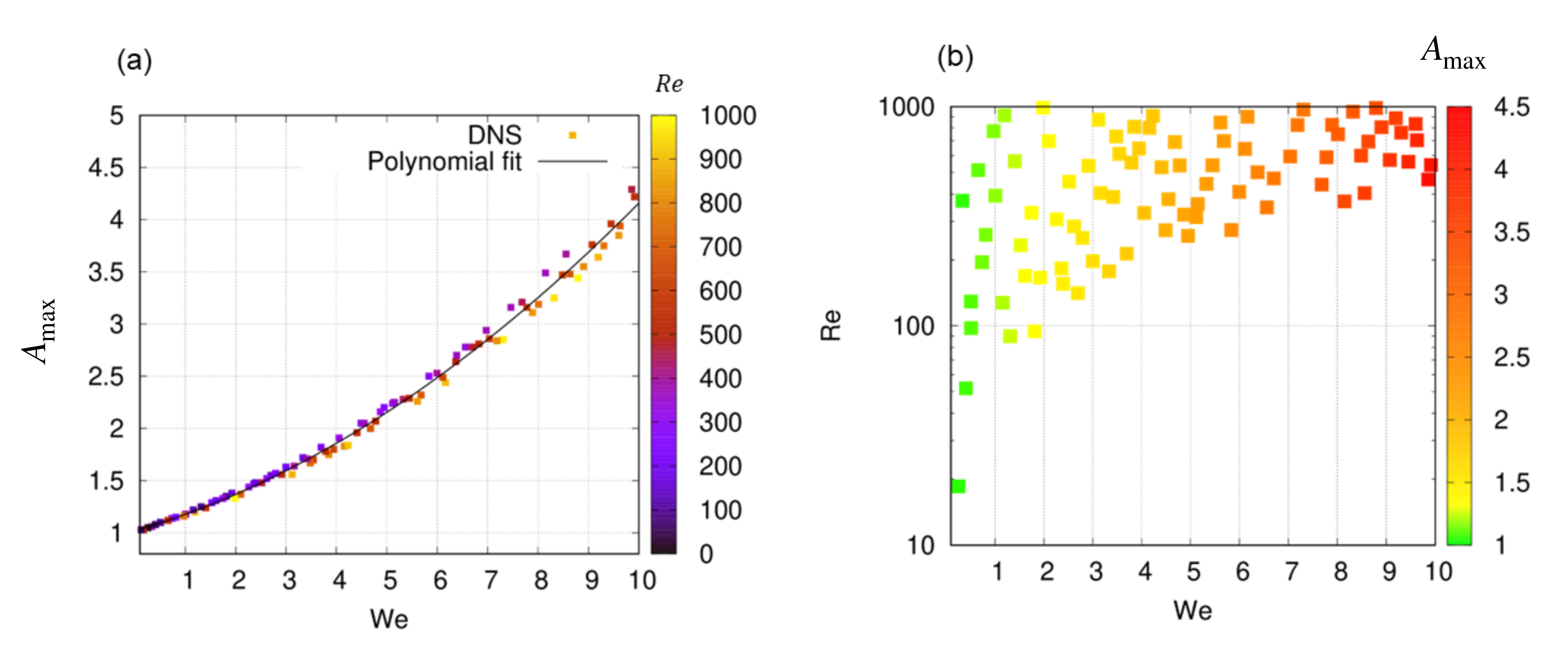}
\caption{(a) Variation of the maximum aspect ratio $A_{\max}=\max(L_y/L_x)$ as a function of \textit{We} for all cases simulated. (b) Variation of the maximum aspect ratio $A_{\max}$ on the \textit{We}-\textit{Re} plane.
}
\label{fig:osc_We_Re_AR}
\vspace{-4mm}
\end{figure}

The \tcr{complex dynamics for drops in finite $\textit{We}$ and $\textit{Re}$ regimes} make it challenging to model drop shape and drag evolutions through conventional physics-based approaches. This complexity is evident in the oscillation amplitude and frequency. Given that oscillation amplitude generally decreases over time due to viscous dissipation, the maximum aspect ratio, $A_{\max}=\max(|L_y/L_x|)$, typically occurring during the first oscillation, is used to represent the oscillation amplitude \cite{Basaran_1992a}. The results for the maximum aspect ratio $A_{\max}$ across all simulated cases are summarized in Fig.~\ref{fig:osc_We_Re_AR}. It is observed that $A_{\max}$ primarily varies with $\textit{We}$. The variation of $A_{\max}$ with $\textit{We}$ is nonlinear and typically follows a quadratic function, as depicted in Fig.~\ref{fig:osc_We_Re_AR}(a). The dependence of $A_{\max}$ on $\textit{Re}$ is generally weak, as seen in Fig.~\ref{fig:osc_We_Re_AR}(b). 
\tcr{For a given $\textit{We}$, $\textit{Oh}$ is inversely proportional to $\textit{Re}$, so the weak dependence on $\textit{Re}$ also indicates that the effect of $\textit{Oh}$ on the oscillation amplitude is small for the ranges of parameters considered. }
%\tcr{For high $\textit{We}$, the differences in $A_{\max}$ for different $\textit{Re}$ become slightly more profound. Drops with higher \textit{Re} generally show lower oscillation amplitude $A_{\max}$ due to the reduced viscous effect. The cases with higher \textit{Re} also exhibit a lower \textit{Oh}, which also contributes to the slight reduction of $A_{\max}$.}

\begin{figure}
\centering
\includegraphics[trim={0cm 0cm 0cm 0cm},clip,width=0.95\textwidth]{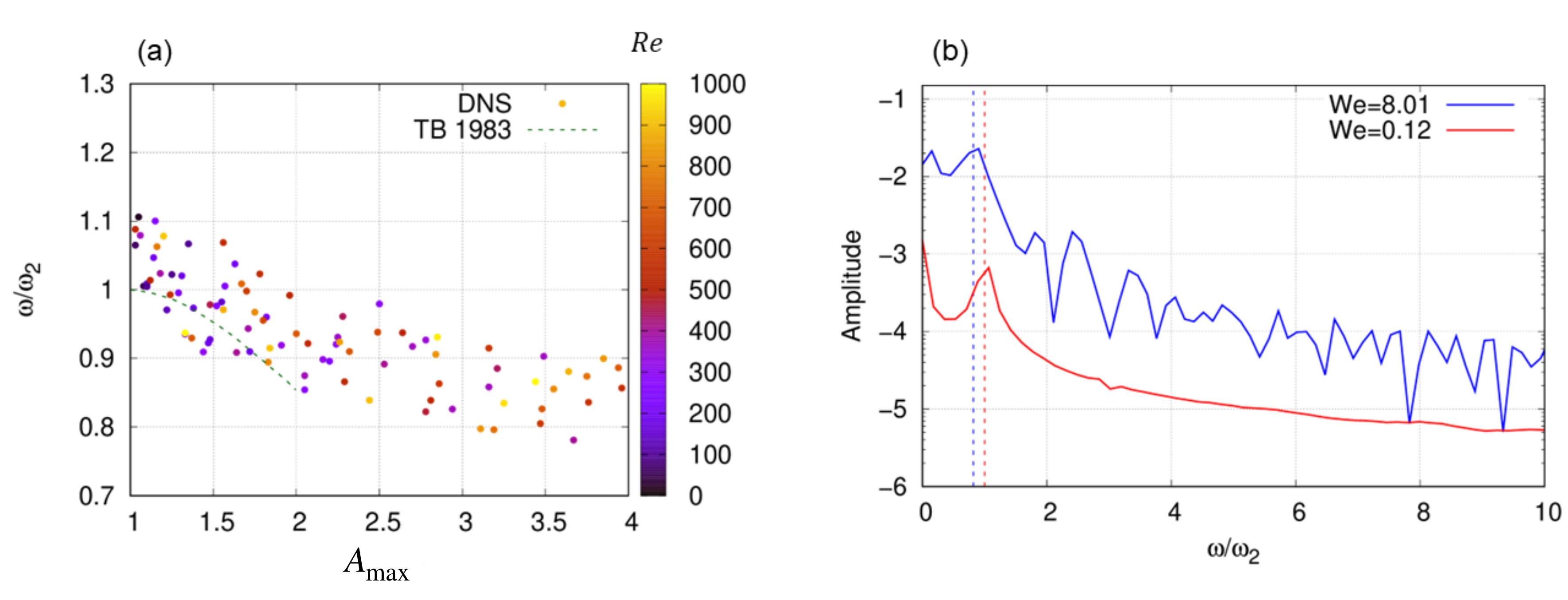}
\caption{(a) Variation of dominant frequency with maximum aspect ratio $A_{\max}=\max(L_y/L_x)$. The color represents 
\textit{Re}. (b) Frequency spectra for two cases with different \textit{We}. The dashed vertical lines indicate the frequency measured based on the first oscillation. 
}
\label{fig:osc_freq_AR}
\vspace{-4mm}
\end{figure}

While multiple modes with different frequencies are present, the dominant mode is the second mode for all cases, as depicted in the spectra for two representative cases shown in Fig.~\ref{fig:osc_freq_AR}(b). The normalized dominant frequency $\omega/\omega_{2,Lamb}$ is plotted as a function of $A_{\max}$ and $\textit{Re}$ in Fig.~\ref{fig:osc_freq_AR}(a). A decrease in frequency is observed as $A_{\max}$ increases, attributable to the increase in $\textit{We}$, as shown in Fig.~\ref{fig:osc_We_Re_AR}(a). The dispersion of data points is due to the limited spectral resolution to identify the dominant frequency. Improving spectral resolution would require running simulations for a much longer time, which is beyond the scope of this study. Nonetheless, a general decreasing trend of $\omega/\omega_{2,Lamb}$ with $\textit{We}$ or $A_{\max}$ is evident. This decrease in frequency, due to nonlinear effects in large-amplitude oscillation, aligns reasonably well with the Tsamopoulos and Brown (TB) nonlinear model when $A_{\max}$ is moderate \cite{Tsamopoulos_1983a}. However, as $A_{\max}$ increases, significant deviations from the TB model are observed, as also observed in previous numerical studies \cite{Basaran_1992a}.

The uncertainties in identifying the dominant frequency render the results somewhat noisy, making it challenging to clearly demonstrate the impact of $\textit{Re}$ on the dominant frequency. For high $\textit{We}$ cases, oscillation frequency changes over time; as the oscillation amplitude decreases, the nonlinear effect on frequency is reduced. Consequently, the frequency during the initial oscillation may slightly differ from that reflected in the spectrum, which accounts for all oscillations, as shown in Fig.~\ref{fig:osc_freq_AR}(b).

In summary, these results demonstrates the difficulty of predicting drop shape oscillation through physics-based approaches, highlighting the necessity for data-driven modeling.

\section{Machine-learning data-driven models}
\label{sec:model}
\subsection{Model overview}
The purpose of the present data-driven model is to predict the temporal evolutions of the drop shape and the drag based on the initial flow conditions and the fluid properties, characterized by the dimensionless parameters like \textit{We} and \textit{Re}. Such a model can be used in Lagrangian spray simulations to predict the motion and shape of individual drops without solving the flows on the drop scale. As described above, the drop shape is characterized by the modal coefficients $C_{0} - C_{10}$ and the drop drag is represented by the drag coefficient. In the simulations, the above quantities are calculated and collected at a fixed timestep \tcr{$\Delta t_{model}\sqrt{\sigma/\rho_l D_0^3}=0.01$} for model training and testing. The simulations have been run to \tcr{$t_{sim}\omega_{2,Lamb}=10.17$ (equivalent to $t_{sim}\sqrt{\sigma/\rho_l D_0^3}=8.00$)} in general. As a result, there are in total 800 \tcr{temporal datapoints} of modal and drag coefficients for one sample (\ie, simulation case). For some cases with $\textit{We}\ge7$, the drop will leave the domain before the specified end time, then for those cases the simulations were stopped as the drop reaches a distance of $2D_0$ from the right boundary. There are 24 cases in the training set \tcr{that} fall into this category and the minimum simulation time is $t_{sim}\omega_{2,Lamb}=6.43$. For those cases, we will have smaller numbers of \tcr{datapoints}. \tcr{Note that the timestep $\Delta t_{\text{model}}$ in the dataset for model development is larger than the timestep $\Delta t_{\text{sim}}$ used in the simulation, and the downsampling is done to provide a compressed dataset without aliasing errors.}

\begin{figure}
\centering
\includegraphics[trim={0cm 0cm 0cm 0cm},clip,width=1\textwidth]{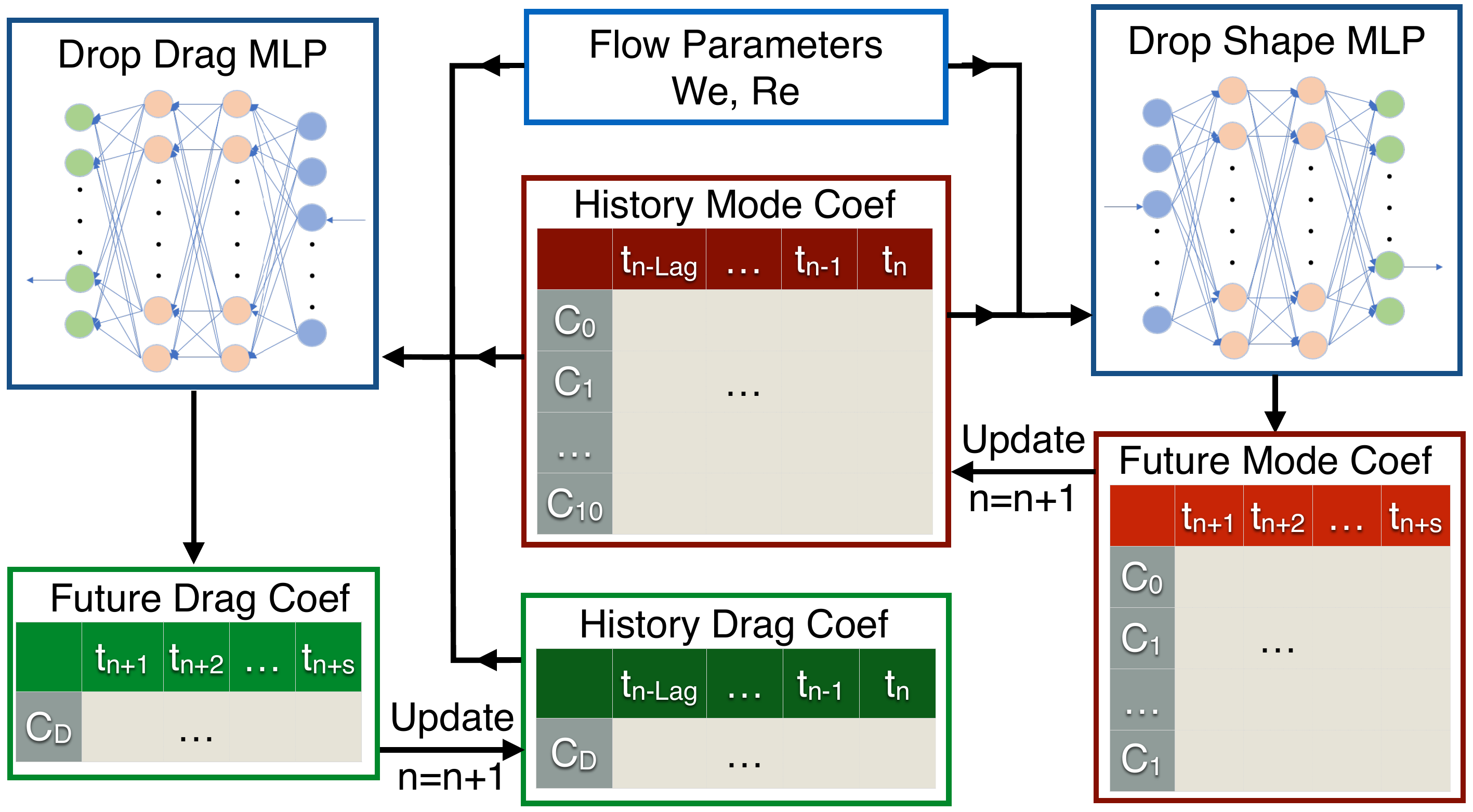}
\caption{Overall modeling strategy.}
\label{fig:osc_model_flowchart}
\vspace{-4mm}
\end{figure}

The overall machine-learning model is depicted in Fig.~\ref{fig:osc_model_flowchart}. The model comprises two multi-layer perceptron (MLP) models. The first model is the drop shape model, which predicts future modal coefficients based on the history of modal coefficients and the dimensionless parameters $We$ and $Re$. The second model is the drop drag model, which uses the history of both modal and drag coefficients, $We$ and $Re$ as inputs to predict future drag coefficients. The rationale for separating the models is to reduce input redundancy, as the history of $C_D$ is not required for predicting the future shape of the drop. Thus, the first model exclusively uses the history of modal coefficients.

Normalization of data to minimize differences across various cases enhances training efficiency. The modal coefficients defining the drop shape are normalized by $\textit{We}$, i.e., $C_n^* = {C_n}/{\textit{We}}$, because the amplitude variation of $C_n$ increases with $\textit{We}$. Similarly, the drag coefficient is normalized by the standard drag for the corresponding $\textit{Re}$, i.e., $C_D^* = C_D / C_{D,std}$.

\subsection{Nonlinear Auto-Regressive models with eXogenous inputs(NARX) recurrent neural network }
\label{sec:NARXNN}
The {Nonlinear Auto-Regressive with eXogenous inputs (NARX) recurrent neural network is a machine-learning model that predicts the system's outputs based on a nonlinear combination of previous outputs and exogenous input parameters \cite{Lin_1995a, Lin_1996a}. This model has demonstrated its accuracy in modeling complex, time-dependent systems, as indicated in prior research \cite{Haris_2022a, Menezes-Jr_2008a, Benouar_2023a}. In the present study, the physical parameters $\textit{Re}$ and $\textit{We}$ serve as the exogenous inputs. The NARX model can be expressed as a mapping of future predictions of the vector of quantities of interest $\mathbf{y}=(C_0, C_1, ... C_{10}, C_D)$ from the history data and exogenous inputs, i.e., 
\begin{align}
\textbf{y}(t_n,\ldots,t_{n+1},t_{n+s-1})=F[\textbf{y}(t_{n-1}),\ldots,\textbf{y}(t_{n-\textit{lag}}),\textit{Re},\textit{We}]
\label{eq:NARXNN_model}
\end{align}
where \textit{stride} ($s$) and $\textit{lag}$ are model hyperparameters that represent the number of future and history timesteps used in each prediction, respectively.

\subsection{Model training and testing}
A NARX neural network can be open-loop or closed-loop. In the present model, the training process is open-loop, for which the simulation data from the history steps are used as inputs to predict the future steps. The open-loop training process for the drop shape model is shown in Fig.~\ref{fig:osc_NARX_architecture}(a). Here, $\textit{lag}=50$ and $s=10$ were used, which means that the modal coefficients from the preceding 50 timesteps are utilized to predict the values for the subsequent 10 timesteps. Consequently, one input ``dataset" to the drop shape MLP consists of 552 inputs, including the 11 modal coefficients for the 50 history steps, plus $\textit{We}$ and $\textit{Re}$, and the MLP output consists of 110 outputs, specifically the modal coefficients for the future 10 steps. Examples for the input-output pairs of two subsequent datasets are depicted in Fig.~\ref{fig:osc_NARX_architecture}(a). For the dataset using modal coefficients for $t_{n-50}, t_{n-49}..., t_{n}$ as inputs, the corresponding outputs for the future steps are $t_{n+1}, t_{n+2}..., t_{n+10}$, while the subsequent dataset uses the modal coefficients for $t_{n-49}, t_{n-48}..., t_{n}, t_{n+1}$ to predict the values for $t_{n+2}, t_{n+3}..., t_{n+11}$. At time zero, there is no ``history" available. Then synthetic data for $t<0$ are created, where the modal coefficients are taken to be zero, i.e., $C_{0,1,...10}(t<0)=0$, and the normalized drag coefficient is set to one, i.e., $C_D/C_{D,std}(t<0)=1$, according to the initial spherical shape of the drop. As a result, for one sample in the training set, there will be 800 datasets for the 800 timesteps.

The data for all samples with different combinations of $\textit{We}$ and $\textit{Re}$ will be combined to the input and output tensors. Then the paired input and output tensors will be shuffled and separated into batches for training. The loss function is the mean squared error (MSE) between model predictions and the ground truth values for the output tensor for one batch. The final loss for an epoch is the average loss across all batches. The MLP comprises 8 hidden layers. Details about the neurons in each layer and other hyperparameters such as learning rate, and batch size are provided in Table \ref{tab:Model_summary}. Early stopping based on evaluating the validation dataset is used to prevent overfitting of the model on training data. 

The testing of the trained model will be closed loop, see Fig.~\ref{fig:osc_NARX_architecture}(b). The predicted outputs for the future 10 timesteps will be fed back as inputs to the model. For example, the previous predicted modal coefficients for $t_{n-50}, t_{n-49}..., t_{n}$ will be used to predict the future values at $t_{n+1}, t_{n+2}..., t_{n+10}$, which will then be used as the next input dataset to the MLP, \ie, the modal coefficients for $t_{n-50}, t_{n-39}..., t_{n+10}$, to predict the values for $t_{n+11}, t_{n+12}..., t_{n+20}$. Similar subsequent predictions will be made. Similar to the training process, synthetic data for $t<0$ will be used to initiate the model prediction. Therefore, the present model is recurrent and will only require $\textit{Re}$ and $\textit{We}$ to autonomously predict the temporal evolutions of the modal coefficients. 

\begin{figure}
\centering
\includegraphics[trim={0cm 0cm 0cm 0cm},clip,width=1.\textwidth]{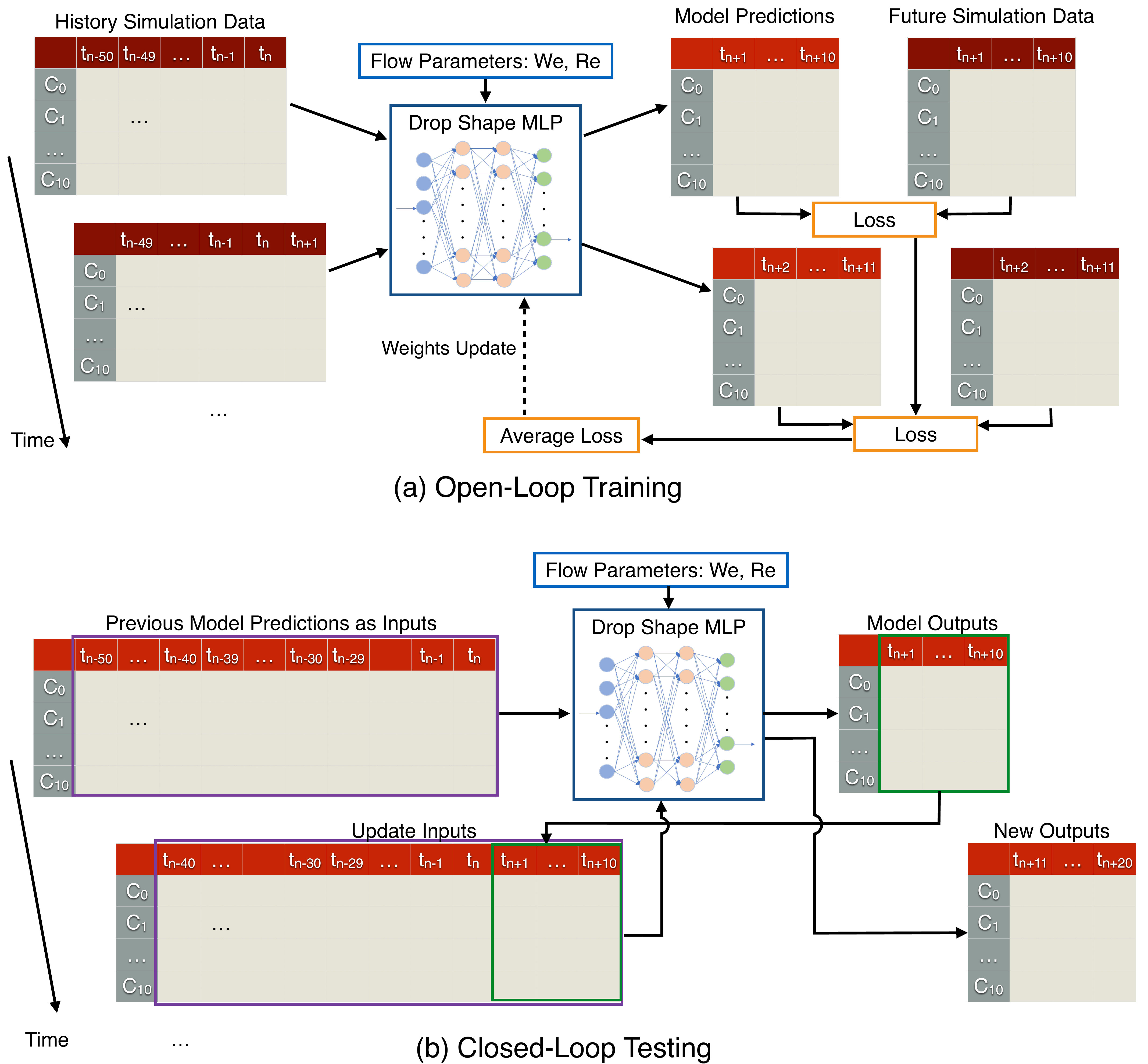}
\caption{Schematic flowchart to show the (a) open-loop training and (b) closed-loop testing processes for the drop shape model.}
\label{fig:osc_NARX_architecture}
\vspace{-4mm}
\end{figure}

\begin{table}
\setlength{\tabcolsep}{3.5pt}
  \caption{Parameters for the NARX neural network model.}
  \centering
  \begin{tabular}{cccc} 
  \hline
  \textit{Hyperparameters} &\textit{Value} \\ 
  \hline
  \textit{Total Number of Inputs for One Dataset for Drop Shape MLP} & $552$ \\
  \textit{Total Number of Inputs for One Dataset for Drop Drag MLP} & $602$ \\
  \textit{Total Number of Outputs for One Dataset for Drop Shape MLP} & $110$ \\
  \textit{Total Number of Outputs for One Dataset for Drop Drag MLP} & $10$ \\
  \textit{Number of Hidden Layers} & $8$ \\
  \textit{Total Number of Neurons in Hidden Layers} & $4450$ \\
  \textit{Batch Size} & $16$ \\
  \textit{Learning Rate} & $10^{-5}$ \\
  \textit{Weight Decay} & $10^{-7}$ \\ 
  \textit{Lag} & $50$ \\
  \textit{Stride,} $s$& $10$ \\
  \textit{Maximum Epoch Number} & $1000$ \\
  \textit{Optimizer} & $\textit{AdamW}$ \\
  \textit{Loss Function} & $\textit{Mean square error}$ \\
  \hline
  \end{tabular}
  \label{tab:Model_summary}
\vspace{-4mm}
\end{table}

The drop drag model architecture is similar the shape model. The MLP for the drop drag has the same number of hidden layers and neurons as the drop shape MLP. The only difference is that, the time history data of $C_d^*$ were added to the inputs. As a result, the drop drag model takes historical values of modal and drag coefficients as temporal inputs and $\textit{We}$ and $\textit{Re}$ as exogenous inputs, to predict future values $C_d^*$. In total, one dataset for the drop drag MLP has 602 inputs and 10 outputs, see table \ref{tab:Model_summary}. Similar to the shape model, drag model also uses open-loop data flow during training and closed loop for testing. 

\tcr{
The average simulation time for one case is about 165 core-hours using an AMD EPYC 7002 processor. The average training times per epoch for the drop shape and drag MLP networks are about 19.32 and 16.03 seconds, respectively, using an NVIDIA RTX 2070 GPU. The number of epochs varies with cases, and the maximum number of epochs is 1000. Once the model is trained, it takes 0.45-0.58 seconds to predict the time evolutions of modal and drag coefficients.
}

\subsection{Model evaluation}
The model performance will be evaluated using error metrics, including the Mean Relative Errors (MRE) for the modal coefficients ($\textit{C}_{MRE}$) and drag coefficients ($\textit{C}_{d_{MRE}}$). As the modal coefficients enable reconstruction of the drop shape, specifically $R(\theta)$ in spherical coordinates, we also assess the MRE for the drop radius ($R_{MRE}$). The error expressions are as follows:
\begin{align}
    \textit{C}_{MRE} & = \frac{1}{N}{\sum_{n=1}^{N}\frac{||{\hat{\textbf{C}}-\textbf{C}||_{F}}}{||\textbf{C}||_{F}}}
\label{eq:osc_coeff_error}\\
    \textit{C}_{d_{MRE}} & = \frac{1}{N}{\sum_{n=1}^{N}\frac{||{\hat{\textbf{C}_{d}}-\textbf{C}_{d}||_{2}}}{||\textbf{C}_{d}^*||_{2}}}
\label{eq:osc_cd_error}\\
    R_{MRE} & = \frac{1}{N}{\sum_{n=1}^{N}\frac{||{\hat{\textbf{R}}-\textbf{R}||_{F}}}{||\textbf{R}||_{F}}}
\label{eq:osc_radius_error}
\end{align}
where $N$ denotes the total number of data points evaluated. Variables with and without $\hat{}$ represent ground truth (simulation data) and model predictions, respectively. The matrix $\textbf{C}$ includes the modal coefficients and all timesteps and samples, while $\textbf{R}$ and $\textbf{C}_{d}$ are vectors for the drop radius at various colatitudes and drag coefficient at different timesteps for all samples. Subscripts $F$ and $2$ denote Frobenius and $L_2$ norm errors for matrices and vectors, respectively.

\section{Data-driven model predictions}
\label{sec:model_results}
\subsection{Model prediction for drop shape deformation}
\label{sec:drop_def}
The model predictions of the modal coefficients ($C_0$-$C_{10}$) for two cases in the testing dataset, i.e., $(\textit{We},\textit{Re})=(0.89,116.12)$ and $(8.31, 652.63)$, representing low and high \textit{We} regimes in the parameter space, are shown in Figs.~\ref{fig:osc_NARX_pred_mode_01} and \ref{fig:osc_NARX_pred_mode_06}, respectively. It is noteworthy that these two test cases were never ``seen" by the model. The predicted time evolutions of all the modal coefficients align remarkably well with the simulation results. The high-order modal coefficients exhibit complicated oscillations, which are accurately captured by the model. Though some discrepancies were observed at later times for the higher-order modes, like $C_5$-$C_{10}$, their amplitudes are small compared to the dominant mode $C_2$. As a result, the discrepancies have a negligible effect on the overall drop shape. Figure \ref{fig:osc_NARX_pred_shape} shows the drop shapes at different times reconstructed from the modal coefficients, illustrating that the predictions for both cases closely match the simulation results. This excellent agreement confirms the model's ability to accurately predict the drop shape evolutions for different \textit{We} and \textit{Re}.

\begin{figure}
\centering
\includegraphics[trim={0cm 0cm 0cm 0cm},clip,width=1\textwidth]{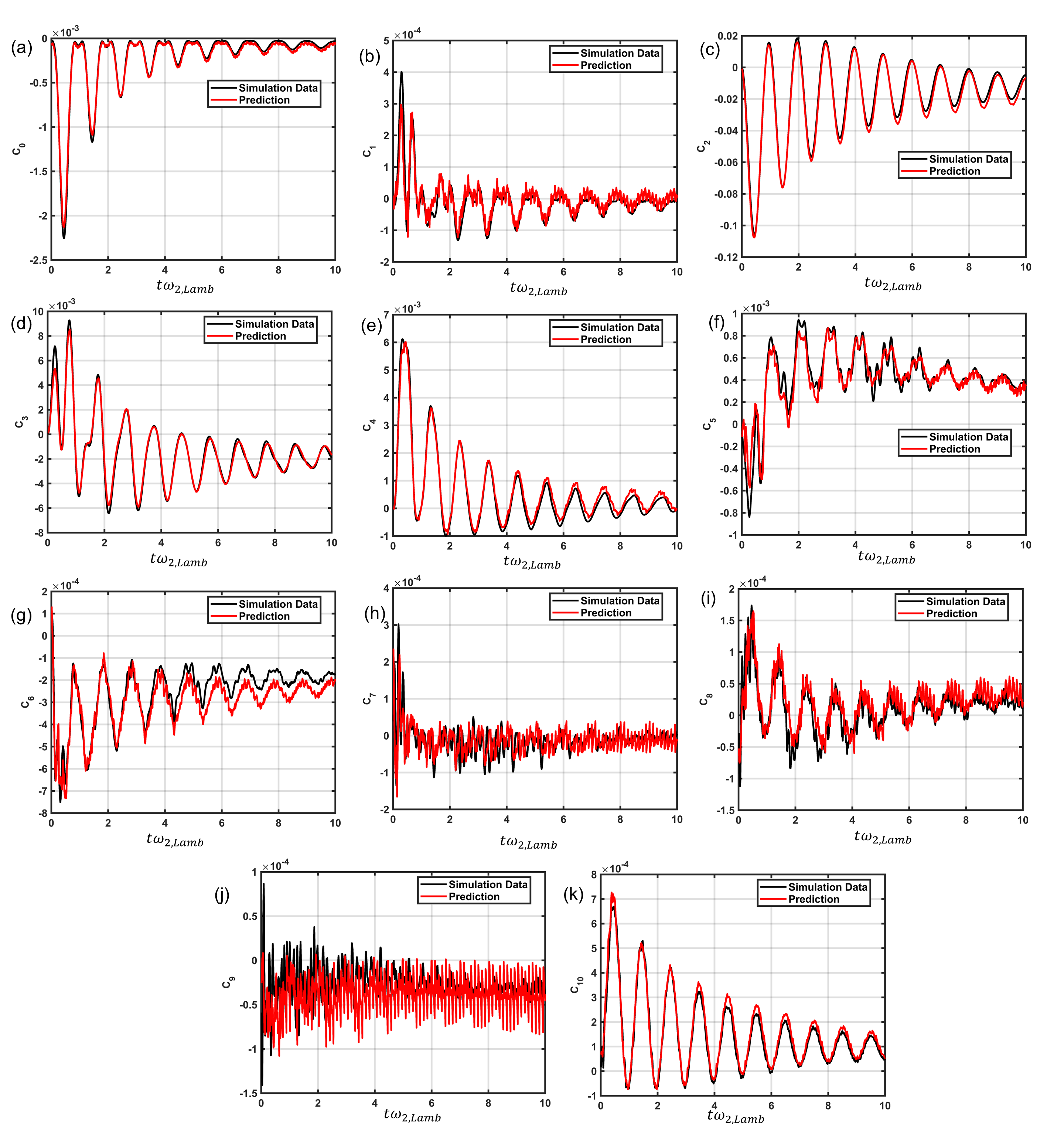}
\caption{Comparison between model predictions of modal coefficients ($C_0$-$C_{10}$) and simulation data for the test case ($\textit{Re}=116.12$, $\textit{We}=0.89$)}
\label{fig:osc_NARX_pred_mode_01}
\vspace{-4mm}
\end{figure}

\begin{figure}
\centering
\includegraphics[trim={0cm 0cm 0cm 0cm},clip,width=1\textwidth]{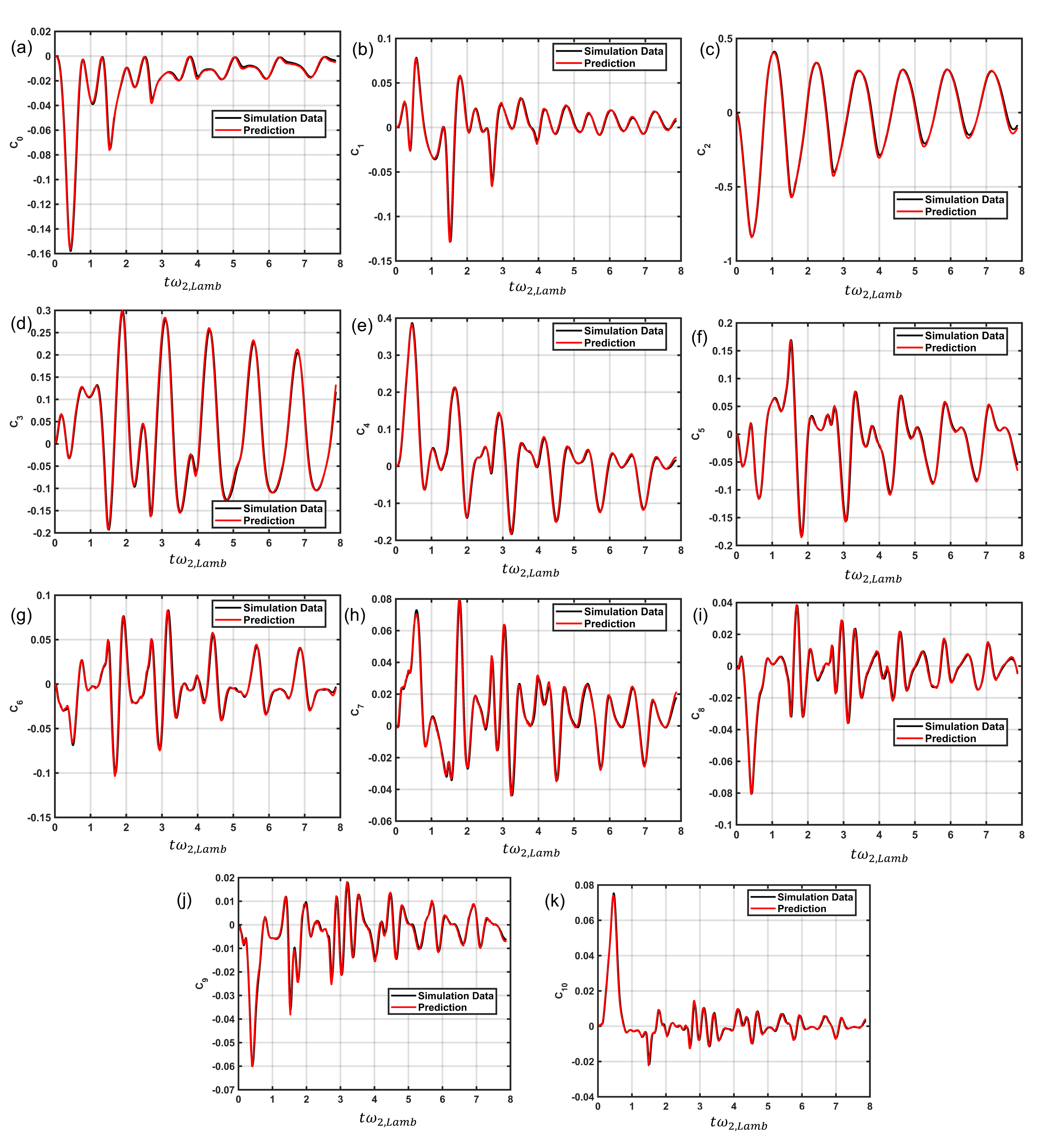}
\caption{Comparison between model predictions of modal coefficients ($C_0$-$C_{10}$) and simulation data for the test case ($\textit{Re}=652.63$, $\textit{We}=8.31$)}
\label{fig:osc_NARX_pred_mode_06}
\vspace{-4mm}
\end{figure}

\begin{figure}
\centering
\includegraphics[trim={0cm 0cm 0cm 0cm},clip,width=1\textwidth]{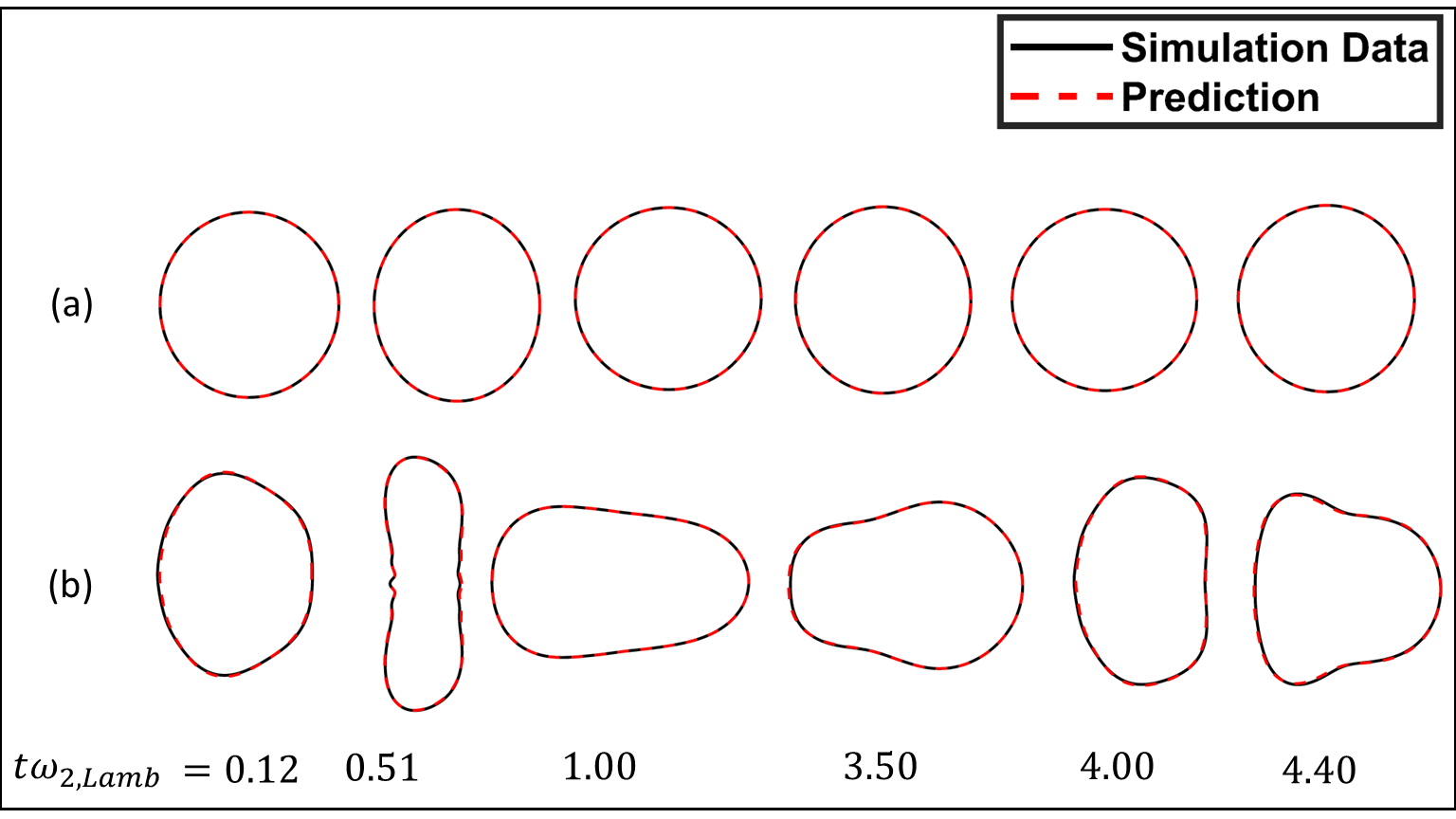}
\caption{Comparison between model predictions of drop shapes and simulation data for the test cases (a) $\textit{Re}=116.12$, $\textit{We}=0.89$, and  (b) $\textit{Re}=652.63$, $\textit{We}=8.31$.}
\label{fig:osc_NARX_pred_shape}
\vspace{-4mm}
\end{figure}

Table~\ref{tab:Osc_dataset_error} summarizes the radius and modal coefficients prediction errors for training, validation and testing datasets. The MRE in modal coefficients ($\textit{C}_{MRE}$), around $8-9\%$, may seem high. However, the high-order modes, which contribute significantly to this error, see Fig.~\ref{fig:osc_NARX_pred_mode_01}, have minor impacts on the drop shape. Thus, the relatively higher errors in the high-order modal coefficients do not significantly affect the accuracy of the predicted drop shape, which is reflected by the MRE for the drop radius ($R_{MRE}$). It is observed that $R_{MRE}$ values are less than $1\%$ for all datasets.

\begin{table}
\setlength{\tabcolsep}{3.5pt}
  \caption{Summary of MRE result for different datasets.}
  \centering
  \begin{tabular}{ccccc} 
  \hline
  \textbf{Datasets} & \textbf{${R}_{MRE}$} & \textbf{${C}_{MRE}$} & \textbf{${C}_{d_{MRE}}^*$}\\
  \hline
  \textit{Training Sets} & $0.80\%$ & $9.62\%$ & $1.28\%$\\
  \textit{Validation Sets} & $0.84\%$ & $8.36\%$  & $1.30\%$\\
  \textit{Testing Sets} & $0.78\%$ & $8.91\%$  & $0.88\%$\\
  \hline
  \end{tabular}
  \label{tab:Osc_dataset_error}
\vspace{-4mm}
\end{table}

The values of $R_{MRE}$ for all 10 cases in the testing dataset are given in Table~\ref{tab:Osc_test_error}. It can be observed that $R_{MRE}$ increases with $\textit{We}$. The highest error, $1.59\%$, is associated with the case with $\textit{Re}=652.63$ and $\textit{We}=8.307$. As $\textit{We}$ increases, the drop deforms more significantly and the drop shape is also more complex. The magnitude of the modal coefficients also increases with $\textit{We}$. As a result, the error in modal coefficients has a greater influence on the drop shape for cases with higher $\textit{We}$.

\begin{table}
\setlength{\tabcolsep}{3.5pt}
  \caption{MRE results for all cases in the test set.}
  \centering
  \begin{tabular}{cccccc} 
  \hline
  Cases & \textit{Re} & \textit{We}  & ${R}_{MRE}$ & ${C}_{MRE}$ & ${C}_{d_{MRE}}^*$ \\
  \hline
  \textit{1} & $116.13$ & $0.898$ & $0.16\%$ & $9.02\%$ & $0.47\%$\\
  \textit{2} & $364.52$ & $1.307$ & $0.26\%$ & $8.44\%$ & $0.35\%$\\
  \textit{3} & $351.40$ & $2.705$ & $0.42\%$ & $8.09\%$ & $0.42\%$\\
  \textit{4} & $274.80$ & $3.117$ & $0.35\%$ & $7.19\%$ & $0.28\%$\\
  \textit{5} & $472.41$ & $3.979$ & $0.65\%$ & $8.32\%$ & $0.62\%$\\
  \textit{6} & $444.77$ & $5.099$ & $0.80\%$ & $9.14\%$ & $0.73\%$\\
  \textit{7} & $486.27$ & $5.714$ & $1.11\%$ & $10.91\%$ & $1.10\%$\\
  \textit{8} & $501.24$ & $6.499$ & $1.08\%$ & $9.27\%$ & $1.12\%$\\
  \textit{9} & $577.14$ & $7.741$ & $1.34\%$ & $9.19\%$ & $1.73\%$\\
  \textit{10} & $652.63$ & $8.307$ & $1.59\%$ & $9.55\%$ & $1.99\%$\\
  \hline
  \textit{Mean} &  &  & $0.78\%$ & $8.91\%$  & $0.88\%$\\
  \hline
  \end{tabular}
  \label{tab:Osc_test_error}
\vspace{-4mm}
\end{table}

%
%\begin{figure}
%\centering
%\includegraphics[trim={0cm 0cm 0cm 0cm},clip,width=0.7\textwidth]{Error.png}
%\caption{Variation of prediction accuracy with the change of \textit{We} }
%\label{fig:osc_pred_error}
%\vspace{-4mm}
%\end{figure}

\begin{figure}
\centering
\includegraphics[trim={0cm 0cm 0cm 0cm},clip,width=1\textwidth]{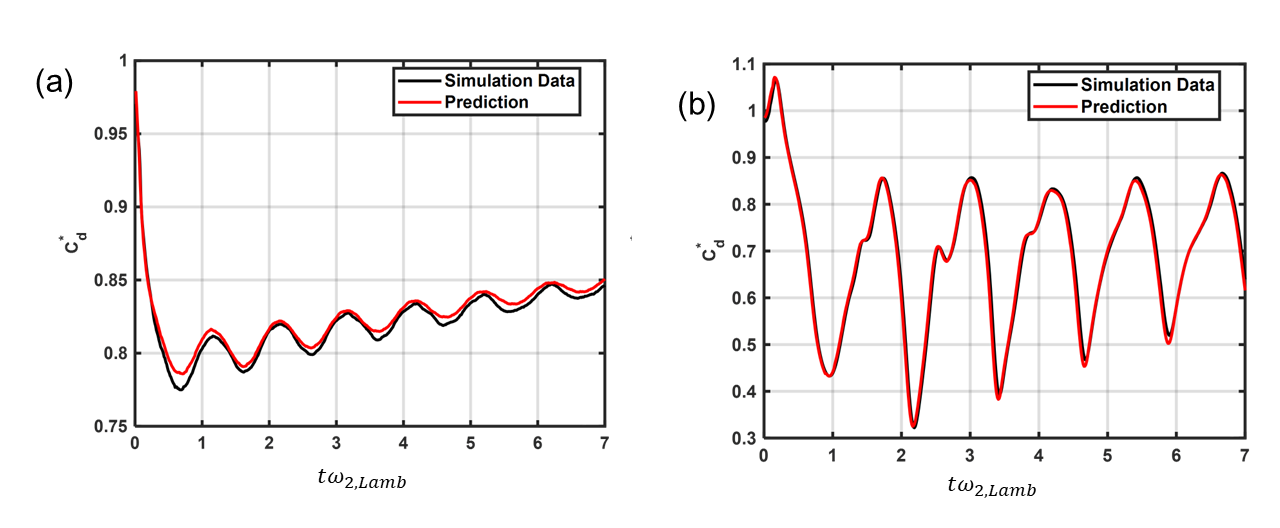}
\caption{Comparison between model predictions of $C_d^*$ and simulation data  for test cases (a) $\textit{We}=0.89$, $\textit{Re}=116$ and (b) $\textit{We}=8.31$, $\textit{Re}=652.63$. }
\label{fig:osc_NARX_pred_drag_test01_06}
\vspace{-4mm}
\end{figure}

\subsection{Model prediction for drop drag coefficient}
With the drop shape model validated, the predictions of the modal coefficients are now incorporated into the drop drag model to predict the time evolution of the drag coefficient. Figure \ref{fig:osc_NARX_pred_drag_test01_06} shows the time evolutions of $C_d^*$ predicted by the present model for two test cases, $(\textit{We},\textit{Re})=(0.89,116.12)$ and $(8.31, 652.63)$. Again, the model predictions agree very well with the simulation data for both cases for all time. Due to the drop shape oscillation, $C_d^*$  evolves in an oscillatory manner, and the model well captures the frequency and magnitude of $C_d^*$.  

The prediction errors of $C_d^*$ for different datasets are shown in Table \ref{tab:Osc_dataset_error}. Due to the error propagation from the shape model to the drag model, the error in drag coefficient is slightly higher than that for the drop radius, but ${C}_{d_{MRE}}^*$ remains lower than 1\% for the testing set. The values of ${C}_{d_{MRE}}^*$ for all cases in the testing set can be found in  Table \ref{tab:Osc_test_error}. Similar to the error in the drop shape, the error in the drag increases with $\textit{We}$, and the values of ${C}_{d_{MRE}}^*$ and ${R}_{MRE}$ are quite similar. The maximum error of $C_d^*$ is 1.99\%, which is associated with the case with  $\textit{Re} = 652.63$ and $\textit{We} = 8.307$. The similarity between ${C}_{d_{MRE}}^*$ and ${R}_{MRE}$ affirms that accurate prediction the drop shape evolution is necessary to well predict the drag and drop motion. 

\section{Conclusions}
\label{sec:conclusions}
A data-driven model has been developed in the present study to predict the shape and drag evolutions of a freely-moving drop in a uniform stream. Assuming the drop fluid and the surrounding gas are water and air, with the drop initially spherical, the key parameters are the Weber (\textit{We}) and Reynolds (\textit{Re}) numbers, defined based on the initial relative velocity and the drop diameter. The present study focuses on the subcritical Weber number regime, in which drops will deform but not break. The significant deformation of the drop influences the interaction between the drop and the surrounding gas flow. As a result, accurately predicting the drag and the resulting motion of the drop requires rigorous prediction of the drop shape evolutions. The complex interplay between drop shape deformation and drag makes conventional physics-based models difficult, so a data-driven approach based on Non-linear Auto-Regressive with eXogenous inputs (NARX) recurrent neural network is adopted.

To provide data for model development, parametric interface-resolved 2D-axisymmetric simulations were performed for 102 cases of different combinations of \textit{We} and \textit{Re} numbers in the parameter space of interest. The geometric volume-of-fluid (VOF) method is used to resolve the sharp interface, and an adaptive quadtree mesh with the minimum cell size equivalent to 128 cells across the initial drop diameter is used in the simulation. To characterize the instantaneous drop shapes, the drop radius as a function of the colatitude is decomposed into spherical harmonic modes. Assuming the drop is axisymmetric, only the axisymmetric modes are considered. The amplitudes of the modal coefficients decrease with the mode number, and a truncation is made at the mode number 10. Eventually, the temporal data for the modal and drag coefficients are collected from the simulations. The modal coefficients are normalized by \textit{We} and the drag coefficient is normalized by the standard drag for spherical particles to reduces the differences across cases. 

The overall model consists of two NARX models. The first model, the drop shape model, takes the history of modal coefficients as temporal inputs, along with \textit{We} and \textit{Re} as exogenous inputs, and predicts the modal coefficients at future timesteps. The second model, the drop drag model, uses the history of modal and drag coefficients, along with \textit{We} and \textit{Re}, as inputs and then predicts the drag coefficients in future steps. The overall simulation data is split into training (92 cases) and testing sets (10 cases). The NARX model training is open-loop, meaning the history simulation data are used as inputs, while testing of the model is done in a closed-loop manner, where the predictions are fed back as inputs, so that the model prediction is recurrent. The model accurately predicts the evolution of both the modal and drag coefficients for both low and high \textit{We} cases. The modal coefficients can be used to reconstruct the drop shape, and prediction errors are also measured on the reconstructed drop radius as a function of the colatitude. The drag coefficient evolves in time in an oscillatory manner due to the drop shape oscillation, and the model predictions agree very well with the simulation data for cases in the testing set. The mean relative errors in the predictions of drop radius and drag coefficient generally increase with \textit{We}. The maximum errors in the testing cases are 1.59\% and 1.99\% for drop radius and drag coefficient, respectively.

\begin{appendix}
\tcr{
\section{Parameters for simulation dataset}
\label{sec:dataset}
Figure~\ref{fig:dataset_table} shows the initial drop diameter $D_0$ and the free-stream velocity $U_0$ for all 102 simulation cases, along with the corresponding values of key dimensionless parameters $\textit{We}$, $\textit{Re}$, and $\textit{Oh}$. Other fluid properties are kept constant. The simulation cases are then split into training, validation, and testing datasets, as indicated.
}

\begin{figure}[tbp]
  \centering
   \includegraphics[width=0.99\textwidth, trim = {1cm 4cm 5cm 2.5cm},clip] {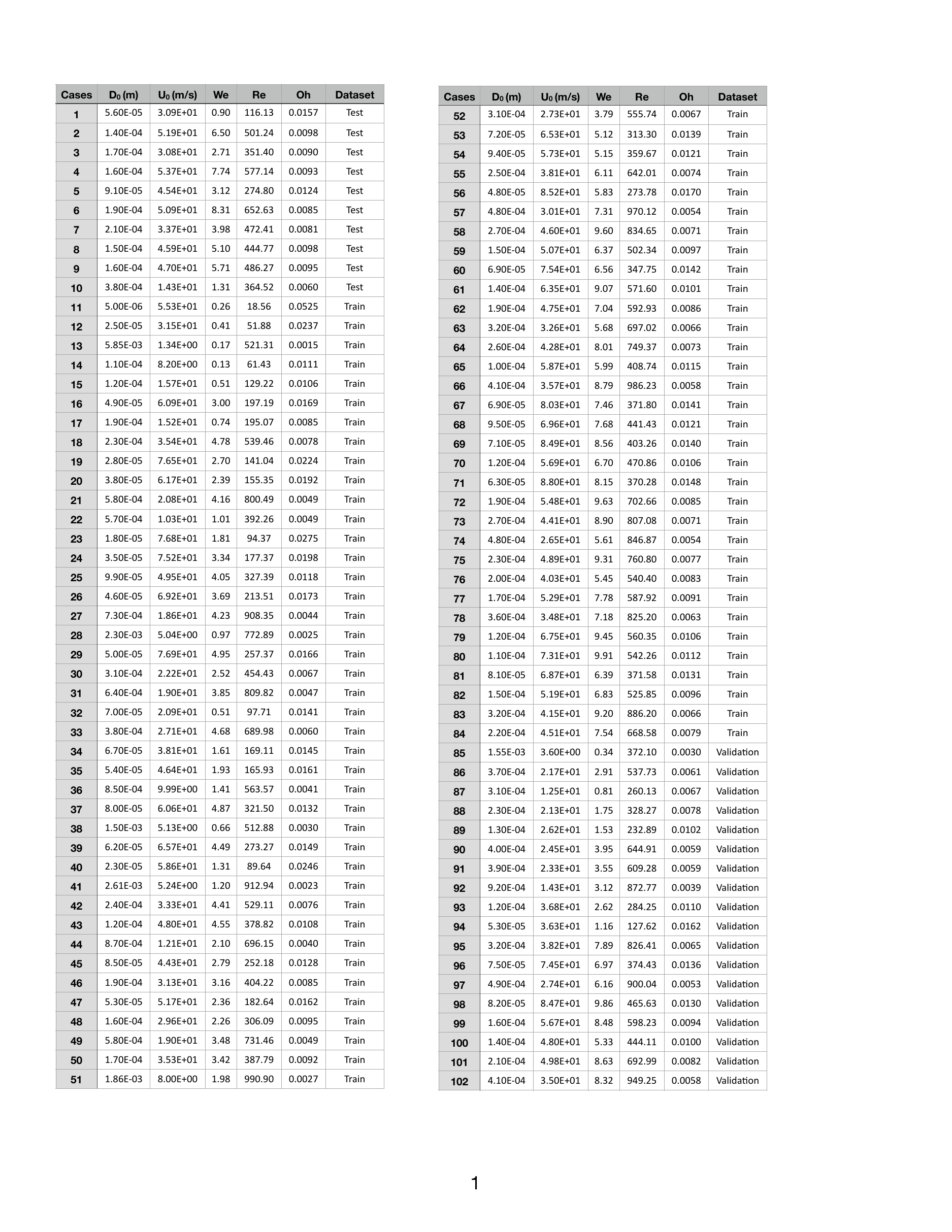}
\caption{Parameters of all simulation cases and corresponding dataset.}
\label{fig:dataset_table}
\end{figure}
\end{appendix}

\section*{Acknowledgments}
This research was supported by ACS-PRF (\#62481-ND9) and NSF (\#1942324). The authors also acknowledge the ACCESS program for providing the computational resources that have contributed to the research results reported in this paper.

\end{document}